\documentclass{ametsocV6.1}
\usepackage{amsmath,amsfonts,amssymb,bm}
\usepackage{mathptmx}%{times}
\usepackage{newtxtext}
\usepackage{newtxmath}
\usepackage{graphicx}
\graphicspath{{plots/}}
\usepackage{lineno}

\nolinenumbers
\title{Estimating the Kinetic Energy Spectrum from the Second-Order Velocity Structure Function using a Regularized Fitting Approach}

\authors{Ayantika Bhattacharjee,\aff{a}\correspondingauthor{Ayantika Bhattacharjee, ayantika36@tamu.edu\\  \textit{
This work has been accepted for publication in the Journal of Atmospheric and Oceanic Technology. Copyright in this work may be transferred without further notice.}} 
Spencer Jones,\aff{a} 
Dhruv Balwada,\aff{b} 
Shane Elipot,\aff{c}
Manuel O. Gutierrez-Villanueva,\aff{b}
}
\affiliation{\aff{a}{Texas A\&M University, College Station, TX}\\
\aff{b}{Lamont-Doherty Earth Observatory, Columbia University, Palisades, NY}\\
\aff{c}{Rosenstiel School of Marine, Atmospheric, and Earth Science, University of Miami, Miami, FL}
}

\abstract{Ocean turbulence plays a key role in shaping large-scale circulation, heat uptake, and biogeochemical processes. The kinetic energy (KE) wavenumber spectrum is a fundamental diagnostic, quantifying how KE is distributed across spatial scales. The second-order structure function---computed from velocity differences between spatially separated observations---provides a complementary measure, but unlike the KE spectrum, it reflects a non-local, weighted integral of KE over all scales. Analytic relationships link the two metrics, permitting forward and inverse transformations between them. However, recovering the KE spectrum from the structure function via the inverse relationship is highly sensitive to sampling limitations and numerical discretization errors. Here we propose a regularized approach in which the spectrum is assumed to consist of a finite number of segments with distinct slopes and amplitudes, and the inversion is formulated as an optimization problem. The approach is first validated in an idealized setting; for a number of idealized KE spectra with prescribed sets of spectral slopes and amplitudes, the corresponding structure functions are computed by numerically evaluating the forward relationship. These structure functions are then used to determine the underlying parameters using our proposed approach, which shows that we are able to perfectly recover the parameters and consequently the KE spectra. The method is further evaluated on high-resolution ocean model output, where it reconstructs the underlying spectra well even in the presence of noise. Finally, we apply the method to surface drifter observations (GLAD and LASER experiments). The results show that the framework enables estimation of the KE spectrum from sparse Lagrangian data, extending spectral diagnostics beyond gridded Eulerian measurements.} %DIF > 
\RequirePackage[normalem]{ulem} %DIF PREAMBLE
\RequirePackage{color}\definecolor{RED}{rgb}{1,0,0}\definecolor{BLUE}{rgb}{0,0,1} %DIF PREAMBLE
\providecommand{\DIFaddbegin}{} %DIF PREAMBLE
\providecommand{\DIFaddend}{} %DIF PREAMBLE
\providecommand{\DIFdelbegin}{} %DIF PREAMBLE
\providecommand{\DIFdelend}{} %DIF PREAMBLE
\providecommand{\DIFaddbeginFL}{} %DIF PREAMBLE
\providecommand{\DIFaddendFL}{} %DIF PREAMBLE
\providecommand{\DIFdelbeginFL}{} %DIF PREAMBLE
\providecommand{\DIFdelendFL}{} %DIF PREAMBLE
\providecommand{\DIFscaledelfig}{0.5}
\RequirePackage{settobox} %DIF PREAMBLE
\RequirePackage{letltxmacro} %DIF PREAMBLE
\newsavebox{\DIFdelgraphicsbox} %DIF PREAMBLE
\newlength{\DIFdelgraphicswidth} %DIF PREAMBLE
\newlength{\DIFdelgraphicsheight} %DIF PREAMBLE
\LetLtxMacro{\DIFOincludegraphics}{\includegraphics} %DIF PREAMBLE
\providecommand{\DIFaddincludegraphics}[2][]{{\color{blue}\fbox{\DIFOincludegraphics[#1]{#2}}}} %DIF PREAMBLE
\providecommand{\DIFdelincludegraphics}[2][]{% %DIF PREAMBLE
\sbox{\DIFdelgraphicsbox}{\DIFOincludegraphics[#1]{#2}}% %DIF PREAMBLE
\settoboxwidth{\DIFdelgraphicswidth}{\DIFdelgraphicsbox} %DIF PREAMBLE
\settoboxtotalheight{\DIFdelgraphicsheight}{\DIFdelgraphicsbox} %DIF PREAMBLE
\scalebox{\DIFscaledelfig}{% %DIF PREAMBLE
\parbox[b]{\DIFdelgraphicswidth}{\usebox{\DIFdelgraphicsbox}\\[-\baselineskip] \rule{\DIFdelgraphicswidth}{0em}}\llap{\resizebox{\DIFdelgraphicswidth}{\DIFdelgraphicsheight}{% %DIF PREAMBLE
\setlength{\unitlength}{\DIFdelgraphicswidth}% %DIF PREAMBLE
\begin{picture}(1,1)% %DIF PREAMBLE
\thicklines\linethickness{2pt} %DIF PREAMBLE
{\color[rgb]{1,0,0}\put(0,0){\framebox(1,1){}}}% %DIF PREAMBLE
{\color[rgb]{1,0,0}\put(0,0){\line( 1,1){1}}}% %DIF PREAMBLE
{\color[rgb]{1,0,0}\put(0,1){\line(1,-1){1}}}% %DIF PREAMBLE
\end{picture}% %DIF PREAMBLE
}\hspace*{3pt}}} %DIF PREAMBLE
} %DIF PREAMBLE
\LetLtxMacro{\DIFOaddbegin}{\DIFaddbegin} %DIF PREAMBLE
\LetLtxMacro{\DIFOaddend}{\DIFaddend} %DIF PREAMBLE
\LetLtxMacro{\DIFOdelbegin}{\DIFdelbegin} %DIF PREAMBLE
\LetLtxMacro{\DIFOdelend}{\DIFdelend} %DIF PREAMBLE
\DeclareRobustCommand{\DIFaddbegin}{\DIFOaddbegin \let\includegraphics\DIFaddincludegraphics} %DIF PREAMBLE
\DeclareRobustCommand{\DIFaddend}{\DIFOaddend \let\includegraphics\DIFOincludegraphics} %DIF PREAMBLE
\DeclareRobustCommand{\DIFdelbegin}{\DIFOdelbegin \let\includegraphics\DIFdelincludegraphics} %DIF PREAMBLE
\DeclareRobustCommand{\DIFdelend}{\DIFOaddend \let\includegraphics\DIFOincludegraphics} %DIF PREAMBLE
\LetLtxMacro{\DIFOaddbeginFL}{\DIFaddbeginFL} %DIF PREAMBLE
\LetLtxMacro{\DIFOaddendFL}{\DIFaddendFL} %DIF PREAMBLE
\LetLtxMacro{\DIFOdelbeginFL}{\DIFdelbeginFL} %DIF PREAMBLE
\LetLtxMacro{\DIFOdelendFL}{\DIFdelendFL} %DIF PREAMBLE
\DeclareRobustCommand{\DIFaddbeginFL}{\DIFOaddbeginFL \let\includegraphics\DIFaddincludegraphics} %DIF PREAMBLE
\DeclareRobustCommand{\DIFaddendFL}{\DIFOaddendFL \let\includegraphics\DIFOincludegraphics} %DIF PREAMBLE
\DeclareRobustCommand{\DIFdelbeginFL}{\DIFOdelbeginFL \let\includegraphics\DIFdelincludegraphics} %DIF PREAMBLE
\DeclareRobustCommand{\DIFdelendFL}{\DIFOaddendFL \let\includegraphics\DIFOincludegraphics} %DIF PREAMBLE
\makeatletter %DIF PREAMBLE
\let\sout@orig\sout %DIF PREAMBLE
\renewcommand{\sout}[1]{\ifmmode\text{\sout@orig{\ensuremath{#1}}}\else\sout@orig{#1}\fi} %DIF PREAMBLE
\makeatother %DIF PREAMBLE
\begin{document}
%% Necessary!
\maketitle

%%%%%%%%%%%%%%%%%%%%%%%%%%%%%%%%%%%%%%%%%%%%%%%%%%%%%%%%%%%%%%%%%%%%%
% SIGNIFICANCE STATEMENT/CAPSULE SUMMARY
%%%%%%%%%%%%%%%%%%%%%%%%%%%%%%%%%%%%%%%%%%%%%%%%%%%%%%%%%%%%%%%%%%%%%
%
% If you are including an optional significance statement for a journal article or a required capsule summary for BAMS 
% (see www.ametsoc.org/ams/index.cfm/publications/authors/journal-and-bams-authors/formatting-and-manuscript-components for details), 
% please apply the necessary command as shown below:
%
% Significance Statement (all journals except BAMS)
%
%\statement
%	 Enter significance statement here, no more than 120 words. See \url{www.ametsoc.org/index.cfm/ams/publications/author-information/significance-statements/} for details.
%
%% Capsule (BAMS only)
%%
%\capsule
%       Enter BAMS capsule here, no more than 30 words. See \url{www.ametsoc.org/index.cfm/ams/publications/author-information/formatting-and-manuscript-components/#capsule} for details.
%
%% * * If using twocol mode, you will need to use the commands "twocolsig" and "twocolcapsule" in place of "sig" and "capsule"
%%      to ensure that the text box correctly spans across both columns.
%

%%%%%%%%%%%%%%%%%%%%%%%%%%%%%%%%%%%%%%%%%%%%%%%%%%%%%%%%%%%%%%%%%%%%%
% MAIN BODY OF PAPER
%%%%%%%%%%%%%%%%%%%%%%%%%%%%%%%%%%%%%%%%%%%%%%%%%%%%%%%%%%%%%%%%%%%%%
%

%% In all cases, if there is only one entry of this type within
%% the higher level heading, use the star form: 
%%
% \section{Section title}
% \subsection*{subsection}
% text...
% \section{Section title}

%vs

\section{Introduction}

Oceanic kinetic energy (KE) is injected at large spatial scales of $O(1000 \mathrm{km})$ by winds, tides, buoyancy forcing, and solar heating, and is ultimately dissipated at molecular scales of $O(1 \mathrm{mm})$ through viscous processes \citep{ferrari2009ocean}. Non-linear interactions across spatial scales drive energy cascades from large to small (forward) and from small to large (inverse) scales \citep{vallis2017, scott2005}. Mesoscale ($O(50\text{–}200\,\mathrm{km})$) motions transport tracers horizontally and extract energy from the background stratification \citep{mcwilliams2008nature, dufour2015role}, while submesoscale ($O(1\text{–}10\,\mathrm{km})$) motions play a central role in transferring heat and tracers between the surface and the deep ocean \citep{elipot2021vertical, balwada2018submesoscale, su2020high, torres2025submesoscale}. At the submesoscale, the flow consists of both balanced motions and inertia–gravity waves, which have distinct dynamical properties and spectral signatures \citep{mcwilliams2016submesoscale, barkan2017stimulated}. Because each of these processes leaves a distinct imprint on how KE is distributed across scales, the KE spectrum, through both its amplitude and slope, provides a fundamental diagnostic of multiscale ocean dynamics  \citep{ferrari2009ocean, callies2013interpreting}.

Estimating the KE spectrum via two-dimensional Fourier transforms requires spatially gridded velocity fields, a condition met by few observational platforms. Satellite altimetry is one of the primary remote sensing platforms for studying ocean turbulence spectra, as it provides global maps of geostrophic velocity derived from gridded sea surface height (SSH) \citep{scott2005, steinberg2022seasonality}. However, conventional altimetry primarily resolves geostrophic mesoscale variability at spatial scales of approximately $100-200$ km. Recent advances, most notably the Surface Water and Ocean Topography (SWOT) mission, have improved the spatial resolution of SSH observations, but the derived velocities still rely on the geostrophic approximation and thus do not represent the ageostrophic component of submesoscale flows.

In-situ platforms, including Lagrangian drifters and shipboard Acoustic Doppler Current Profilers (ADCP) provide data at spatial scales well below the resolution of altimetry, making them well suited for studying mesoscale and submesoscale turbulence. Shipboard ADCP observations provide high-resolution velocity along each ship transects  \citep[e.g.][]{callies2013interpreting}, from which one-dimensional wavenumber spectra can be estimated by applying Fourier transforms along the sampling direction under a frozen-flow (fast-tow) assumption. Such analyses have been conducted in the Gulf Stream \citep{buhler2014wave, callies2013interpreting, callies2015seasonality, caofoxkemper2023}, the western Pacific \citep{qiu2017submesoscale}, Drake Passage \citep{rocha(2016mesoscale(a))}, and the eastern Pacific \citep{chereskin2019characterizing, soares2022transition}. Studies combining shipboard ADCP data with high-resolution modeling have further highlighted the coexistence and scale-dependent partitioning of balanced geostrophic motions and internal gravity waves in the submesoscale range \citep{rocha2016seasonality(b), qiu2017submesoscale, torres2018partitioning, torres2022separating}. Nevertheless, ADCP-based spectra are one-dimensional projections of the underlying two-dimensional flow, and the frozen-flow assumption breaks down for rapidly evolving submesoscale motions. In contrast, drifters sample velocity along their irregular Lagrangian paths, making Fourier-based stimation of wavenumber spectra infeasible. They are, however, well suited to two-point statistics such as the second-order structure function, which quantifies velocity variability as a function of spatial separation and thereby characterizes the distribution of energy across scales \citep{balwada2016scale, lacasce2016estimating, davidson2015turbulence, mccaffrey2015estimates}. Major observational programs --- Global Drifter Program (GDP) \citep{lumpkin2007measuring,elipot2016global}, the Grand LAgrangian Deployment (GLAD) experiment \citep{poje2014submesoscale}, and the LAgrangian Submesoscale ExpeRiment (LASER) \citep{novelli2017biodegradable} --- have produced extensive drifter datasets that support robust diagnosis of small-scale ocean dynamics.

Inferring KE spectra from two-point statistics derived from drifter observations remains challenging. Several studies have inferred spectral characteristics from the statistics of pair-separation growth \citep[e.g.][]{koszalka2009relative, poje2014submesoscale, balwada2021relative}.  Most recently, \cite{qian2025inferring} employed relative dispersion statistics --- in particular  the cumulative inverse separation time (CIST; \citealp{lacasce2022relative}), to infer the KE spectra slope from surface drifter observations in the Gulf of Mexico, highlighting pronounced seasonal regimes. However, their approach relies on the Fokker--Planck framework, which assumes a stationary, homogeneous, and isotropic turbulent velocity field, and the inferred spectra reflect primarily balanced (rotational) flows and exclude contributions from unbalanced (divergent) flows such as inertia-gravity waves and submesoscale fronts. Furthermore, this framework is restricted to power-law spectra and can only distinguish slopes within the local regime ($1 < \alpha < 3$, i.e., between $k^{-1}$ and $k^{-3}$), as slopes steeper than $k^{-3}$ produce indistinguishable exponential pair separation growth, making it impossible to differentiate between them.

A more direct approach uses the exact analytic integral relationship between the second-order structure function and the KE spectrum, thereby offering the possibility of near-global spectral diagnostics from existing drifter datasets \citep{balwada2016scale, lacasce2016estimating, davidson2015turbulence, mccaffrey2015estimates}. However, directly inverting this mathematical relationship is non-trivial. \cite{lacasce2016estimating} showed that recovering the longitudinal KE spectrum from the longitudinal structure function is mathematically feasible; however, this method generates a noisy spectrum unless a sixth-order polynomial is fit to the results. In addition, this method is inaccurate for steep spectra.  Reconstructing the full isotropic spectrum from the total structure function, which has contributions from both longitudinal and transverse components of the flow, is still more ill-conditioned: both forward and inverse transforms involve a Bessel function kernel, and the kernel's oscillatory character makes the solution highly sensitive to noise, finite sampling, and the limited range of observed separations. Even in idealized tests, the inversion fails because of discretization errors and incomplete resolution of large-scale contributions.

These difficulties motivate alternative approaches for estimating the KE spectrum from non-gridded sparse velocity observations. In this study, we propose a regularized fitting framework for estimating the KE spectrum from observed second-order structure function under the assumption of homogeneity, without requiring isotropy. This approach stabilizes the ill-conditioned inversion and enables robust estimation of spectral slopes. In our framework, the KE spectrum is represented as a piecewise function composed of multiple power-law segments over a restricted interval that allows us to regularize the problem. The method is validated using a hierarchy of tests: it is first applied to idealized spectra with prescribed slopes to assess its ability to recover known spectral forms, then evaluated using high-resolution surface velocity fields from the MITgcm llc4320 simulation to examine performance under realistic flow conditions, and finally it is applied to surface drifter observations to demonstrate its effectiveness for real-world, sparsely sampled data. Collectively, these analyses show that the proposed framework provides a practical and physically consistent approach by preserving the exact relationship between the KE spectrum and the second-order structure function while enforcing physically plausible spectral forms. 

The paper is organized as follows. Section~\ref{sec: Two-point statistics} introduces two-point statistics and the second-order structure function, including the key equations used in this study. Section~\ref{sec:method} presents the formulation of the regularized method to estimate the KE spectrum from $SF_2$. Sections~\ref{sec:idealized_Data} and \ref{sec: MITgcm} validate the method using idealized spectra and high resolution model output, respectively. Section~\ref{sec: Drifter_data} demonstrates the application of the framework to sparse surface drifter observations from two targeted experiments in the Gulf of Mexico. Finally, Section~\ref{sec: discussion} summarizes the key findings, discusses the implications and limitations of the proposed approach, and outlines directions for future work.

\section{Two-point statistics and Second-Order Structure function}
\label{sec: Two-point statistics}

Two-point statistics provide a framework for characterizing how quantities measured at two locations separated by a distance $r$ are related, and how this relationship varies across spatial scales. They are especially valuable when working with sparse and irregular observations such as those from drifter data. The second-order velocity structure function, $SF_2(r)$, is a central diagnostic in this framework, defined as the second-order raw moment of velocity difference. For a horizontal velocity field $\mathbf{u}(\mathbf{x})$, the $SF_2(r)$ is expressed as
\begin{equation}
SF_2(r) = \left\langle \left| \delta \mathbf{u}(\mathbf{x},\mathbf{r},t) 
\right|^2 \right\rangle
= \left\langle 
\left| \mathbf{u}(\mathbf{x}+\mathbf{r},t) - \mathbf{u}(\mathbf{x},t) 
\right|^2 
\right\rangle,
\label{eq:sf2_def}
\end{equation}
where $\mathbf{x}$ is the horizontal position, $\mathbf{r} = \mathbf{x}_2 - \mathbf{x}_1$ 
is the separation vector with magnitude $r = |\mathbf{r}|$, $\delta \mathbf{u}$ is the 
velocity difference, and $\langle \cdot \rangle$ denotes an ensemble average. Under the 
assumption of statistical homogeneity, adopted throughout this study, $SF_2$ does not 
depend on absolute position $\mathbf{x}$ or time $t$, and the ensemble average can be approximated by 
averaging over all available pairs separated by a distance $r$, which may be distributed 
across different positions and times. This averaging is also performed over all orientations of $\mathbf{r}$, so that $SF_2$ depends only on the scalar separation $r$. 

The second-order structure function may be decomposed into longitudinal and transverse components,
%
% \begin{equation}
% [SF_2(r)]_{LL} = (\delta u_L)^2, \qquad [SF_2(r)]_{TT} = (\delta u_T)^2 ,
% \end{equation}
\begin{equation}
    SF_2^L(r) = \left\langle |\delta u_L|^2 \right\rangle, \qquad 
    SF_2^T(r) = \left\langle |\delta u_T|^2 \right\rangle.
    \label{eq: sf2_LT}
\end{equation}
The longitudinal velocity difference, $\delta u_L$, is defined as the projection of the velocity difference onto the direction of separation, $\delta u_L = \delta \mathbf{u} \cdot \mathbf{r}/|\mathbf{r}|$. The transverse velocity difference, $\delta u_T$, is defined as the projection of the velocity difference $\delta \mathbf{u}$ onto the unit vector $\mathbf{t}$ perpendicular to the separation vector $\mathbf{r}$ in the horizontal plane, such that $\delta u_T = \delta \mathbf{u} \cdot \mathbf{t}$, where $\mathbf{t} = \hat{\mathbf{z}} \times \hat{\mathbf{r}}$, $\hat{\mathbf{r}} = \mathbf{r}/|\mathbf{r}|$, and $\hat{\mathbf{z}}$ is the upward unit vector normal to the horizontal plane. With this convention, $\mathbf{t}$ corresponds to a counterclockwise rotation of $\hat{\mathbf{r}}$ by $90^\circ$ in the horizontal plane. The total second-order structure function can therefore be written as the sum of the longitudinal and transverse components,
\begin{equation}
% SF_2(r) = [SF_2(r)]_{LL}  + [SF_2(r)]_{TT} .
SF_2(r) = SF_2^L(r)  + SF_2^T(r) .
\label{eq:sf2_cal}
\end{equation}

Under the assumption of stationary and homogeneous two-dimensional turbulence, the second-order structure function is related to the KE spectrum $E(k)$ \citep{bennett1984relative} as,
\begin{equation}
SF_2(r) = 2 \int_0^{\infty} E(k) \left[ 1 - J_0(k r) \right] \, dk ,
\label{eq:sf2_exact}
\end{equation}
where $J_0$ is the zeroth-order Bessel function of the first kind and $k$ is the wavenumber. Here $E(k)$ is defined as the radial KE spectrum, 
\begin{equation}
    E(k) \equiv \int_0^{2\pi} \Phi(k, \theta_k) \, k \, d\theta_k,
    \label{eq: ek}
\end{equation}
where $\Phi(k, \theta_k)$ is the 2D velocity power spectral density, 
$k = |\mathbf{k}| = \sqrt{k_x^2 + k_y^2}$ is the scalar radial wavenumber 
with $k_x$ and $k_y$ being the zonal and meridional wavenumber components, 
and $\theta_k \in [0, 2\pi)$ is the azimuthal angle of the wavenumber 
$k_x = k\cos\theta_k$ and $k_y = k\sin\theta_k$. Although $\Phi$ depends on $\theta_k$, which may reflect anisotropy, $E(k)$ in Eq.~\ref{eq:sf2_exact} represents the energy integrated over all directions at a given wavenumber magnitude. Therefore, isotropy is not a necessary assumption for this relation itself. 

The KE spectrum can be derived from Eq.~\eqref{eq:sf2_exact} (shown in Supplemental Material Section 1) as,
\begin{equation}
E(k) = \int_{0}^{\infty} \frac{SF_2(\infty) - SF_2(r)}{4}\, J_0(k r)\, k r \, dr.
\label{eq:Ek_from_SF2}
\end{equation}

While Eqs.~\eqref{eq:sf2_exact} and \eqref{eq:Ek_from_SF2} provide analytical relationships between the second-order structure function and the KE spectrum, their practical interpretation is often guided by scaling arguments. If the KE spectrum scales as $E(k) \sim k^{-(m+1)}$, then the corresponding second-order structure function scales as $SF_2(r) \sim r^{m}$, provided that $0 \le m+1 \le 3$  \citep{babiano1985structure, bennett1984relative, lindborg2007horizontal}. In classical 2D turbulence theory, for an inverse energy cascade with constant energy flux \(\varepsilon\), the second-order structure function scales as \(SF_2 \sim \varepsilon^{2/3} r^{2/3}\) (Kolmogorov’s \(2/3\) law). In contrast, for a forward enstrophy cascade with enstrophy flux \(\eta\), it scales more steeply as \(SF_2 \sim \eta^{2/3} r^{2}\) \citep{lindborg1999can, lacasce2002turbulence}. These regimes illustrate how \(SF_2(r)\) links scaling behavior in physical space to the corresponding spectral slopes.

The second-order structure function, $SF_2(r)$, can be computed from both gridded velocity fields and sparsely sampled observations such as drifter measurements, making it a practical diagnostic for observational studies. Classical turbulence theory and much of our understanding of energy cascades, however, are formulated in spectral space, motivating efforts to infer $E(k)$ from $SF_2(r)$. In practice, applying Eq.~\eqref{eq:Ek_from_SF2} to observational data is difficult (Supplemental Material Figure S1): observational structure functions are available only over a finite range of separations, are contaminated by noise, and the oscillatory Bessel kernel in the integrand amplifies numerical errors, with accurate evaluation requiring very fine sampling and careful cancellation of oscillatory terms that is often unstable in practice \citep{lacasce2016estimating, cree1993algorithms}. These limitations motivate the alternative approach developed in the next section.

\section{Estimating $\mathbf{E(k)}$ from $\textbf{SF}_\textbf{2}$: a regularized approach}
\label{sec:method}

We assume a parametric form for $E(k)$ that is flexible enough to represent the spectral shapes expected in ocean turbulence, while remaining sufficiently constrained to exclude unrealistic solutions. Because turbulent cascades are commonly characterized by power-law behavior with transitions across dynamical regimes, $E(k)$ is represented as a piecewise function consisting of $S$ power-law segments defined over successive equal intervals in logarithmic wavenumber space (i.e., uniform spacing in $\log k$). Each $i^{\text{th}}$ segment is defined by an amplitude $b_i$ and slope $\alpha_i$ over the interval $[k_{i-1}, k_i]$, and the full spectrum is obtained by summing over all segments,

% , compute the corresponding $SF_2(r)$ using the forward relation, and adjust the parameters until the predicted structure function matches that estimated from the data. In this way, recovering $E(k)$ is reduced to estimating a limited set of parameters, which regularizes the inversion.

%
\begin{equation}
E(k) = \sum_{i=1}^{S}  b_i\, k^{\alpha_i} [H(k-k_{i-1}) - H(k-k_{i})],
\label{eq:parametric_form}
\end{equation}
where $H(\cdot)$ is the Heaviside step function. The term $[H(k - k_{i-1}) - H(k - k_{i})]$ acts as a top-hat function,
\begin{equation}
H(k - k_{i-1}) - H(k - k_{i}) =
\begin{cases}
1, & k_{i-1} \le k < k_i, \\
0, & \text{otherwise},
\end{cases}
\label{eq: heaviside_term}
\end{equation}
restricting each power-law contribution to its corresponding interval and ensuring that $E(k)$ vanishes outside the global range $[k_{\min}, k_{\max}]$. 

Substituting this piecewise spectrum into Eq.~\eqref{eq:sf2_exact} yields integrals that are non-elementary and cannot be expressed in closed form. We therefore adopt an indirect approach: we assume the parametric form for $E(k)$, evaluate the corresponding $SF_2(r)$ using the forward relation, and adjust the parameters until the predicted structure function matches that estimated from the data. In this way, the estimation of $E(k)$ is reduced to fitting a limited set of parameters, which regularizes the inversion.

The structure function, $SF_2^{est}(r)$, is estimated directly from drifter observations or model velocity fields on a logarithmically spaced separation grid of $N$ points. The corresponding wavenumber grid for estimating $E(k)$ is defined as,
\begin{equation}
k = \frac{2\pi}{r},
\label{eq:rk_relation}
\end{equation}
which yields a logarithmically spaced $k$ grid of the same size. With this definition, if $r$ is expressed in km, then $k$ is expressed in rad\,km$^{-1}$. For consistency with standard spectral representations, wavenumbers shown in the remainder of this paper are converted to units of cycles km\(^{-1}\), while separation distances are reported in km. The Eq.~\ref{eq:rk_relation} is used solely to establish the correspondence between spatial separation and wavenumber ranges and does not influence the inversion procedure itself. Other equivalent scale--wavenumber relationships may also be used for this purpose without affecting the formulation of the inversion. The wavenumber grid is therefore set by the physical range of separations available in the data: the largest separation $r_{\max}$ controls the lowest resolvable wavenumber $k_{\min} = 2\pi / r_{\max}$, and the smallest separation $r_{\min}$ controls the highest wavenumber $k_{\max} = 2\pi / r_{\min}$. For a given initial set of spectral parameters, \(\{b_i,\alpha_i\}\), the piecewise spectrum is substituted into Eq.~\eqref{eq:sf2_exact}, and the integral is evaluated numerically using Simpson’s rule to obtain the corresponding structure function. This computed structure function ($SF_2^{com}$) is then compared with $SF_2^{est}$, and the spectral parameters are determined by non-linear least squares fitting. To further restrict the solution to physically plausible spectra, the amplitude parameters ($b_i$) are constrained to be non-negative, and the spectral slopes ($\alpha_i$) are restricted to the interval $[-4, 1]$. Continuity across adjacent spectral segments is enforced through the boundary condition,
\begin{equation}
b_i (k_i)^{\alpha_i} = b_{i+1} (k_{i})^{\alpha_{i+1}}.
\label{eq:boundary_condition}
\end{equation}
which reduces the number of parameters from $2S$ ($S$ amplitudes and $S$ slopes) to $S+1$.

The non-linear least-squares fitting is performed using the Trust Region Reflective (TRF) algorithm \citep{coleman1996interior, branch1999subspace} as implemented in the SciPy Python package function \texttt{scipy.optimize.least\_squares} \citep{scipy2020}. TRF supports bound constraints on the parameters, and is generally robust for this type of problem. In our tests,  it produced more consistent and physically realistic solutions than the alternative \texttt{dogbox} algorithm  (\cite{voglis2004rectangular}
; results  not shown), which is also available within the same SciPy Python package. At each iteration, the TRF algorithm locally approximates the cost function and updates the parameters within a prescribed trust region, which limits how much the parameters are allowed to change in a single optimization step. The size of the trust-region is adjusted adaptively at each iteration. It is increased when the local quadratic model provides a good prediction of the actual reduction in the cost function, and decreased when the predicted reduction differs substantially from the actual reduction. This approach is robust and efficient for high-dimensional optimization problems involving many parameters. 

The residual $R(r)$ used in the least-square fitting is defined as the relative error between the computed and estimated $SF_2$, 
\begin{equation}
    R(r) = \Big(\frac{SF_2^{com}(r) - SF_2^{est}(r) }{SF_2^{est}(r)}\Big).
\end{equation}
This choice normalizes the misfit by the magnitude of estimated $SF_2$, preventing scales with larger absolute values from dominating the fit and allowing all separations to contribute uniformly. SciPy provides several loss-function options for non-linear least-squares fitting, but we use the linear loss because alternative choices do not produce noticeable improvement in the results. 

After the least-squares fitting, the best-fit structure function ($SF_2^{fit}$) is obtained, and the error between the $SF_2^{fit}$ and $SF_2^{est}$, defined as,
\begin{equation}
\text{error} = \sqrt{\frac{1}{N}
\sum_{i=1}^{N}
\left(
\frac{SF_2^{\text{fit}}(r_i) - SF_2^{\text{est}}(r_i)}{SF_2^{\text{est}}(r_i)}
\right)^2},
\label{eq:error}
\end{equation}

where $N$ is the number of discrete spatial separation points $r$. The number of power-law segments, $S$, is not known in advance and is therefore determined through the fitting procedure. Therefore, the algorithm is repeated for $S$ ranging from 1 to $N$, with the error between the estimated and best-fitted structure functions computed for each configuration using Eq.~\ref{eq:error}. The configuration of $S$ that minimizes Eq.~\eqref{eq:error} is selected 
as the optimal fit, with the corresponding error and the best-fit structure function referred to as the best-fit error and optimized $SF_2$, respectively. The spectral parameters from the optimal configuration are then used to estimate $E(k)$ on the wavenumber grid derived from Eq.~\eqref{eq:rk_relation}, using Eq.~\eqref{eq:parametric_form}. Idealized tests (not shown) suggest that, for a noise-free $SF_2$, this method of choosing $S$ produces a more accurate $E(k)$ than any other method tested.

\section{Method Validation with Idealized Data}
\label{sec:idealized_Data}

We use idealized KE spectra with prescribed spectral slopes and energy levels to test the accuracy  of the regularized approach. These synthetic spectra mimic those reported from observations and models in the ocean and atmosphere for different dynamical regimes \citep[e.g.,][]{rocha(2016mesoscale(a)), qiu2017submesoscale, callies2015seasonality, lindborg1999can}. In addition, scale-dependent noise is introduced  to mimic more realistic conditions. This controlled framework allows us to directly assess the accuracy and robustness of the method before applying it to realistic oceanic datasets.

\begin{figure}[htbp]
 \centering
 \noindent
 \includegraphics[width=35pc,angle=0]{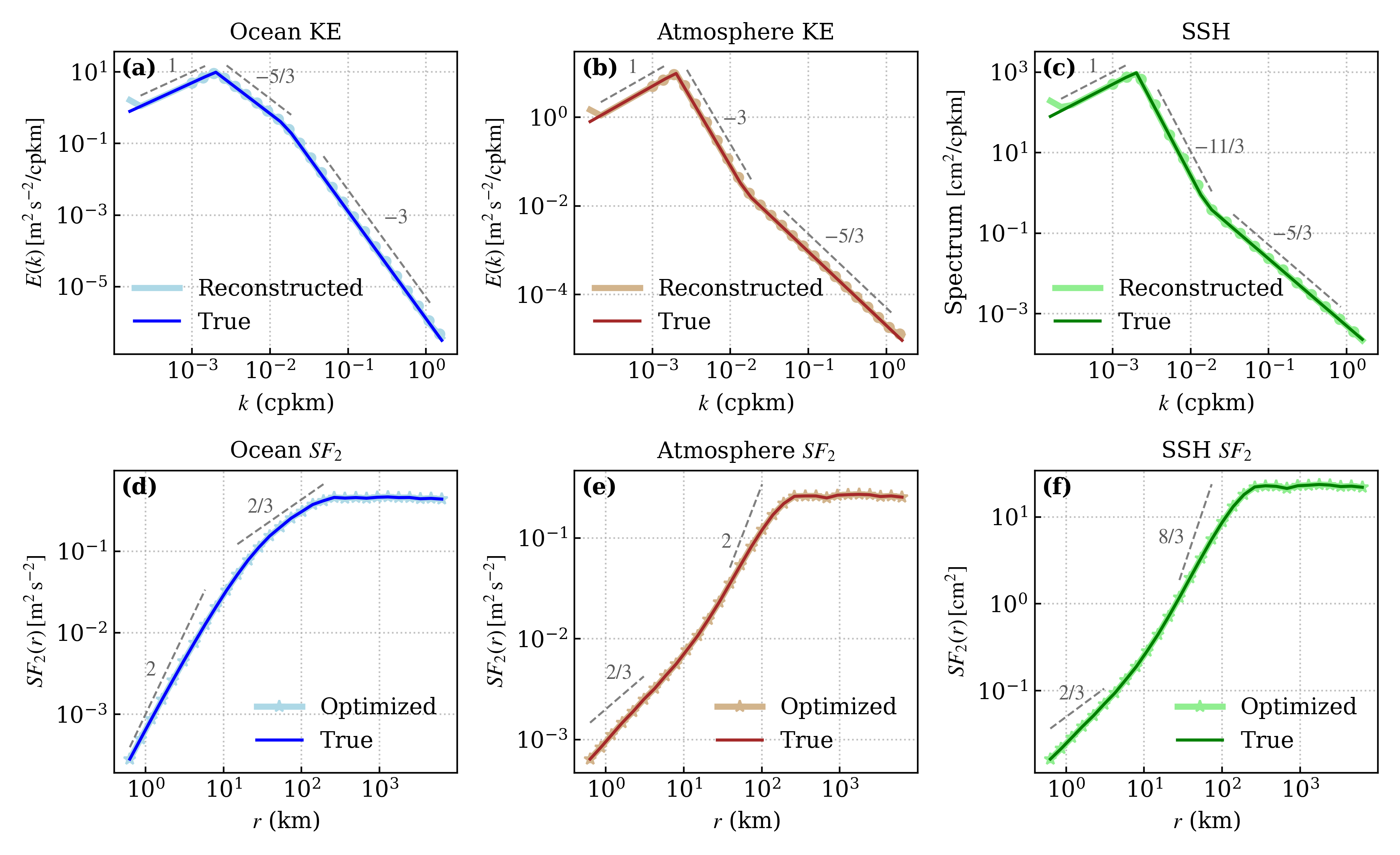}
 \DIFaddendFL \caption{Panels (a)--(c) show the idealized (dark colors) and reconstructed (light colors) wavenumber spectra for ocean KE (blue), atmospheric KE (brown), and SSH (green), respectively. The dots mark the spectral segment boundaries. Panels (d)--(f) show the corresponding true (dark colors) and optimized (light colors) second-order structure functions. The lighter curves nearly overlap the darker ones in all panels. Dashed gray lines indicate reference power-law slopes.}
 \label{fig:ps_ideal_fitting}
\end{figure}

\subsection{Idealized Kinetic Energy Spectra}

We apply our method to a set of idealized spectra constructed to resemble the characteristic spectral behaviors of oceanic and atmospheric KE, and sea surface height (SSH) variability (Figure~\ref{fig:ps_ideal_fitting}(a--c)). For the oceanic KE case (Figure \ref{fig:ps_ideal_fitting}(a), blue curve), we prescribe a spectrum composed of three power-law segments with slopes of $+1$, $-5/3$, and $-3$, consistent with ranges reported in previous studies \citep{lacasce2016estimating, callies2015seasonality}. For atmospheric KE (Figure \ref{fig:ps_ideal_fitting}(b), brown line), we prescribe the idealized spectrum with slopes of $+1$, $-3$, and $-5/3$, following the canonical scaling arguments discussed by \citet{lindborg1999can}. For SSH variability (Figure \ref{fig:ps_ideal_fitting}(c), green line), we consider slopes of $+1$, $-11/3$, and $-5/3$, motivated by recent analyses of SSH wavenumber spectra \citep{wang2025enhanced}. We note that the SSH spectral behavior at higher wavenumbers remains uncertain; therefore, the $-5/3$ slope is used here only as an approximate representation. The $k$ axis is discretized on a logarithmically spaced grid spanning $1.5 \times 10^{-4}$ to $1.5$ cpkm (\(10^{-3}\) to \(10^{1}\) rad/km) with 30 points ($N = 30$), and the $r$-axis, used to estimate $SF_2(r)$, is calculated using Eq. \eqref{eq:rk_relation}. For each idealized spectrum, the associated second-order structure function $SF_2(r)$ is computed at these $r$ values using Eq.~\eqref{eq:sf2_exact} and is shown as the dark solid curves in Figure~\ref{fig:ps_ideal_fitting}(d-f). As discussed in Section \ref{sec: Two-point statistics}, the slope correspondence \(\mathrm{SF}_2 \sim r^{m} \leftrightarrow E(k) \sim k^{-(m+1)}\) is an asymptotic relation valid only for \(0 \le m+1 \le 3\). So, at the lower wavenumber, when the spectral slope approaches 1, which is $m+1 = -1$, the corresponding \(SF_2\) slope approaches zero (Figure~\ref{fig:ps_ideal_fitting}(d-f)). It is important to note that for \(m+1>3\), the \(SF_2\) scaling saturates at \(r^2\). The second-order structure functions, computed from idealized spectra,  are first substituted into Eq.~\ref{eq:Ek_from_SF2}, but the resulting spectra are noisy because of the oscillatory nature of the Bessel function, finite sampling, and the limited range of separation, as shown in Supplemental Material Figure S1. We therefore apply the regularization method described in Section~\ref{sec:method} to reconstruct the corresponding spectra from these second-order structure functions.

\begin{figure}[htbp]
 \centering
 \noindent\includegraphics[width=20pc,angle=0]{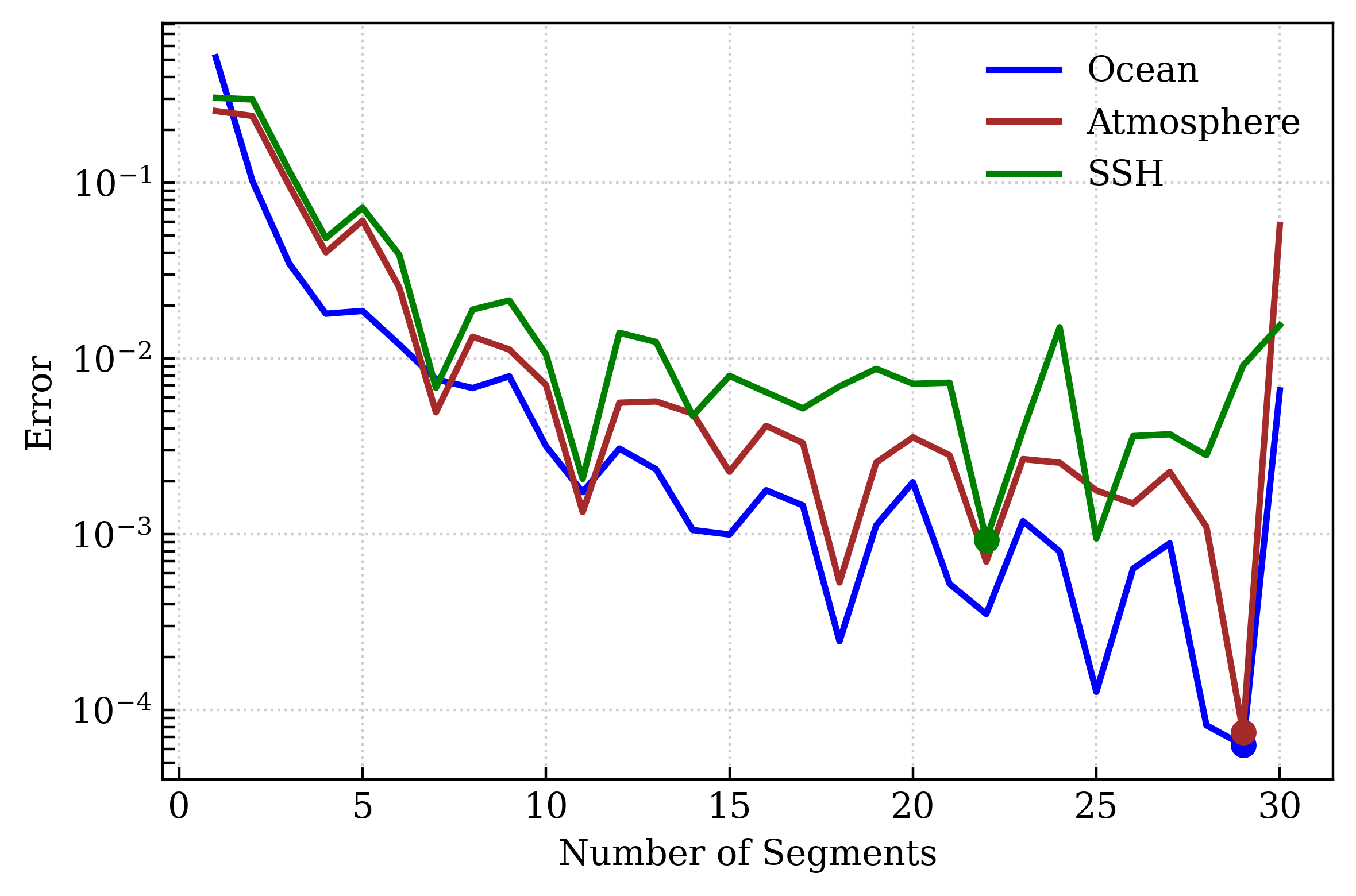}
 \caption{Error between the true and optimized second-order structure functions ($\text{SF}_2$), calculated using Eq.~(\ref{eq:error}), as a function of the number of segments for the ocean, atmosphere, and SSH cases. Dots denote the optimal spectral segment selected for each case, corresponding to the minimum error (best-fit).}
 \label{fig: err_ideal}
\end{figure}\textbf{}

In Figure~\ref{fig:ps_ideal_fitting}(d–f), the optimized structure functions (light colors) obtained using the regularized approach closely  overlap with the true structure functions (dark colors), indicating an excellent fit. Figure~\ref{fig: err_ideal} shows the best-fit error is on the order of $10^{-5}$ for the oceanic and atmospheric cases, and $10^{-4}$ for SSH. The optimal number of spectral segments is 29 for the oceanic and atmospheric cases and 22 for SSH (Figure~\ref{fig: err_ideal}), resulting in $S+1$ parameters, that is, 30 for the oceanic and atmospheric cases and 23 for SSH. The KE spectra, reconstructed using these optimal parameters (Figure~\ref{fig:ps_ideal_fitting}(a–c), light-colored curves), perfectly match the prescribed idealized spectra, demonstrating that the method successfully recovers the spectral structure. 

These results show that the regularized framework accurately recovers both the spectral slopes and amplitudes of the prescribed spectra across all three dynamical regimes, providing a baseline for assessing its performance under more realistic conditions that include noise and more complex flow fields. An advantage of the present framework is that it is not restricted to the $-3$ to $-1$ slope range often associated with classical power-law arguments. Instead, by estimating $E(k)$ through constrained forward fitting of $SF_2(r)$, the method can recover spectral slopes over the broader range from $-4$ to $+1$, as permitted by the imposed bounds on $\alpha$.
\begin{figure}[htbp]
 \centering
 \noindent\includegraphics[width=39pc,angle=0]{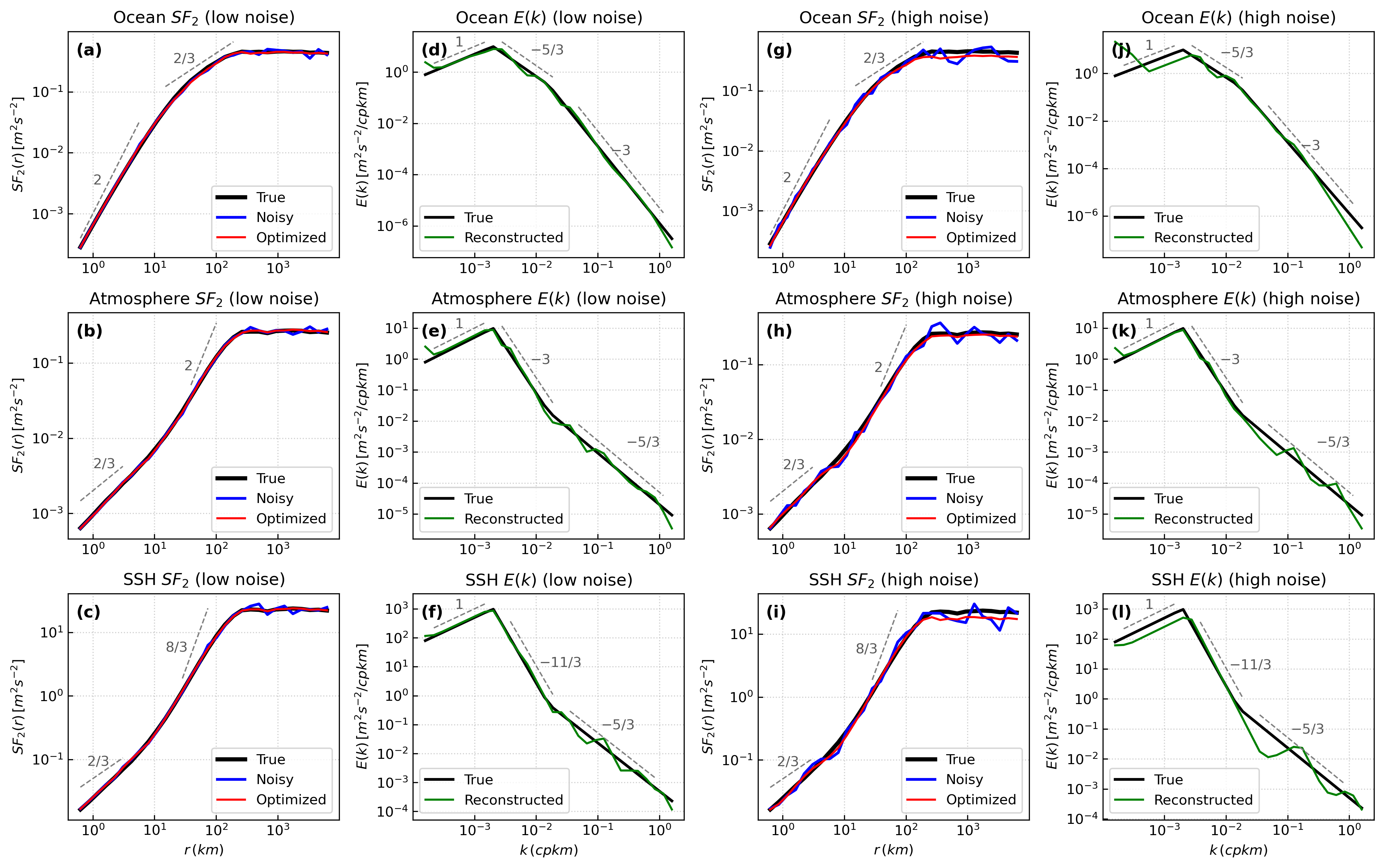}
 \caption{Three-panel comparison of structure function fitting and reconstructed KE spectra for idealized ocean, atmosphere, and SSH cases under low- and high-noise conditions. Left two columns (a--f) show the low-noise case, and right two columns (g--l) show the high-noise case. For each case, scale-dependent multiplicative $\chi^2$ noise is added to $SF_2$, with noise model parameters given in Table~\ref{tab:noise_params}.
}
 \label{fig: sf2_ensemble}
\end{figure}
\subsection{Sensitivity to Scale-Dependent Noise} 

To evaluate the performance of the method under realistic condition, we test the method using noisy synthetic structure functions. In observational data, uncertainties in $SF_2(r)$ are usually not uniform across $r$. For example in data collected through drifter experiments, at small separations, the large number of independent velocity pairs leads to relatively low sampling noise, whereas at larger separations the reduced number of independent samples results in increased uncertainty. This behavior was noticed in GLAD and LASER drifter observations \citep{balwada2022}. To mimic the scale-dependent uncertainty inherent in observational structure functions, we prescribe different noise levels across distinct ranges of spatial separation. Since $SF_2(r)$ is a variance-like quantity, its sampling variability is more appropriately described by a chi-squared ($\chi^2$) distribution under the assumption of Gaussian noise in the velocity field. Accordingly, multiplicative scale-dependent noise is introduced using a chi-squared random variable $X(r)$ drawn from a $\chi^2$ distribution. The perturbation is defined as
\begin{equation}
\epsilon(r) = \sigma(r) \, \frac{X(r) - \mathrm{df(r)}}{\sqrt{2\,\mathrm{df(r)}}},
\end{equation}
where $\mathrm{df}(r)$ is the scale-dependent degrees of freedom that control the statistical shape of the distribution, and $\sigma(r)$ is the scale-dependent noise amplitude that controls the magnitude of the perturbation at each scale $r$. The noisy structure function is then defined as,
\begin{equation}
SF_2^{\mathrm{noisy}}(r) = SF_2(r)\,\bigl(1 + \epsilon(r)\bigr).
\end{equation}

We apply scale-dependent noise perturbations to the second-order structure functions derived from the same idealized spectra shown in Figure~\ref{fig:ps_ideal_fitting}, for all three cases: ocean, atmosphere, and SSH. The noise model parameters are summarised in Table~\ref{tab:noise_params}, with noise amplitudes chosen as representative values broadly consistent with the observational variability typically seen in drifter structure functions. The parameter $df(r)$ sets the shape of the $\chi^2$ distribution from which the noise is drawn: large values produce a near-Gaussian perturbation, while small values produce a more skewed perturbation. The amplitude $\sigma(r)$ sets the relative size of the perturbation as a fraction of $SF_2(r)$. At small separations ($r < 2$~km), $\sigma(r)$ is small, producing a low-amplitude noise that mimics the well-constrained $SF_2(r)$ expected at these scales. At intermediate separations, $\sigma(r)$ is chosen to mimic moderate sampling noise. At large separations ($r \geq 3\times10^{2}$~km), $\sigma(r)$ is increased, yielding a larger-amplitude noise that emulates the rapid loss of independent pair counts that characterizes observational $SF_2(r)$ at large $r$. The two columns of $\sigma$ in Table~\ref{tab:noise_params} define the low- and high-noise scenarios used to assess the robustness of the reconstruction.

\begin{table}[h!]
\centering
\caption{Scale-dependent noise parameters applied to the second-order structure functions. The degrees of freedom $\mathrm{df}(r)$ control the statistical shape of the $\chi^2$ noise distribution, while $\sigma(r)$ sets the noise amplitude as a fraction of the signal at each scale.}
\label{tab:noise_params}
\begin{tabular}{lcccc}
\hline
\textbf{Scale} & \textbf{Range} & $\mathrm{df}(r)$ & $\sigma$ \textbf{(low-noise)} & $\sigma$ \textbf{(high-noise)} \\
\hline
Small        & $r < 2$ km                    & 1000 & 2\%  & 10\% \\
Intermediate & $2 \leq r < 3\times10^{2}$ km & 100  & 5\%  & 15\% \\
Large        & $r \geq 3\times10^{2}$ km     & 15   & 10\% & 25\% \\
\hline
\end{tabular}
\end{table}

In both the low- and high-noise experiments (Figure~\ref{fig: sf2_ensemble}(a--c) and (g--i)), the optimized $SF_2(r)$ (red curves) match the true $SF_2(r)$ (black curves) more closely than the noisy $SF_2(r)$ (blue curves). In the low-noise case, the reconstructed spectra (Figure ~\ref{fig: sf2_ensemble}(d--f), green curves) closely follow the true spectra 
(black curves), although the transition scales at which the spectral slopes change are not precisely recovered. In the high-noise case, the reconstruction is degraded more noticeably (Figure~\ref{fig: sf2_ensemble}(j--l)). At low wavenumbers, the positive slope is recovered but the energy level is underestimated. The intermediate slope is reasonably well recovered for the atmosphere and SSH cases, but less accurately for the ocean case. At higher wavenumbers, the slope is recovered less accurately for all cases. The number of spectral segments selected by the optimization also varies with noise level (not shown): for the ocean case, the optimal number decreases from 22 in the low-noise case to 15 in the high-noise case; for the atmosphere case, it increases from 26 to 29; and for the SSH case, it decreases from 25 to 20. These differences indicate that the optimal model complexity depends on both the noise level and the spectral characteristics of the system. Overall, the method is robust in recovering the broad spectral structure under low noise, but its ability to resolve slope transitions, energy levels, and fine-scale behaviour decreases with increasing noise, as expected.

% Next, we evaluate the accuracy of the methodology to reconstruct the $E(k)$ from $SF_2(r)$ using the horizontal velocity outputs ($u,v$) from the MITgcm llc4320 simulation. This allows us to benchmark the reconstructed $E(k)$ against the true expected isotropic spectra under realistic conditions. 

\section{Method Validation with model Data}
\label{sec: MITgcm}

\begin{figure}[htbp]
 \centering
 \noindent\includegraphics[width=38pc,angle=0]{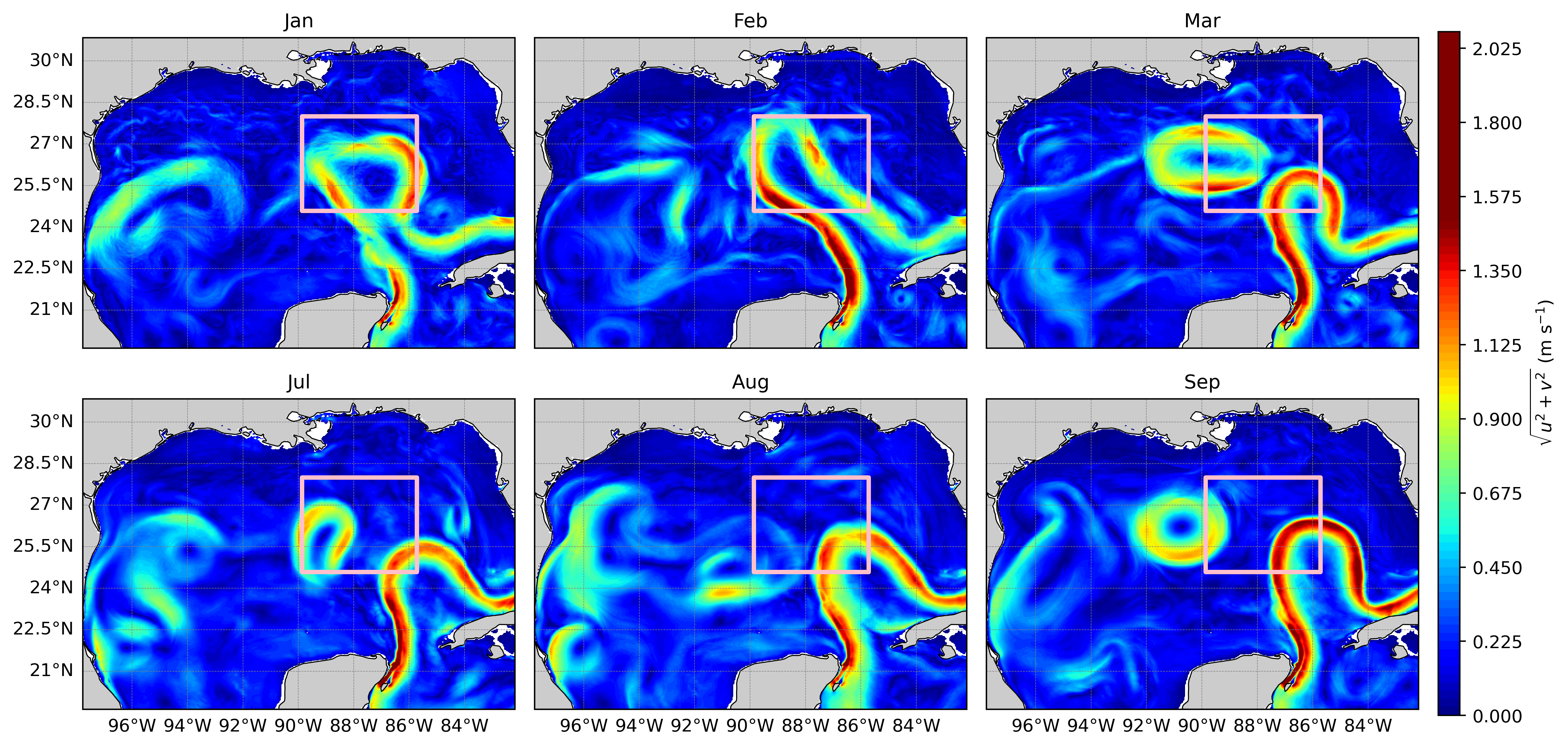}
 \caption{Monthly mean surface velocity from the MITgcm llc4320 for winter (January--March) and summer (July--September). The pink box outlines the $400$ km $\times 400$ km  domain used in this study.}
 \label{fig:domain}
\end{figure}

Having demonstrated good performance in idealized cases, we next evaluate our method using velocity fields from the high-resolution MIT General Circulation Model (MITgcm) llc4320 simulation to assess our approach in a dynamically consistent oceanic setting. The llc4320 configuration uses a global latitude--longitude--cap grid with a horizontal grid spacing of $1/48^{\circ}$ (approximately 2.3 km) and an effective resolution of about 8 km \citep{marshall1997hydrostatic,rocha(2016mesoscale(a))}, fully resolving mesoscale motions and internal tides, and partially resolving submesoscale dynamics. The simulation therefore provides a plausible representation of upper-ocean turbulence across a broad range of scales, making it well suited for testing scale-dependent diagnostics. In this study, a $400~\text{km} \times 400~\text{km}$ subdomain is selected in the Gulf of Mexico (Figure~\ref{fig:domain}), a region characterized by energetic mesoscale and submesoscale activity \citep{qian2025inferring}. We further examine the sensitivity of the reconstructed KE spectrum to seasonal variations in spectral shape, as the Gulf of Mexico exhibits distinct summer and winter dynamical regimes.

\subsection{Structure Function and Spectrum Estimation from Model Velocity Field}

Daily surface velocity snapshots are used to compute second-order structure functions and KE spectra. January--March and July--September are selected to represent winter and summer conditions, respectively, thereby capturing the seasonal contrast in the flow field. For each month, five snapshot at five-day intervals are selected for analysis. Prior to analysis, the velocity fields are spatially detrended to remove large-scale background gradients, and a Tukey window is applied to reduce spectral leakage associated with the finite spatial domain. For consistency, the same windowing procedure is used in the calculation of both the KE spectrum, and the second-order structure function, $SF_2(r)$. These quantities are then computed for each selected day, and the resulting estimates are averaged to obtain a monthly mean.

\begin{figure}[htbp]
 \centering
 \noindent\includegraphics[width=32pc,angle=0]{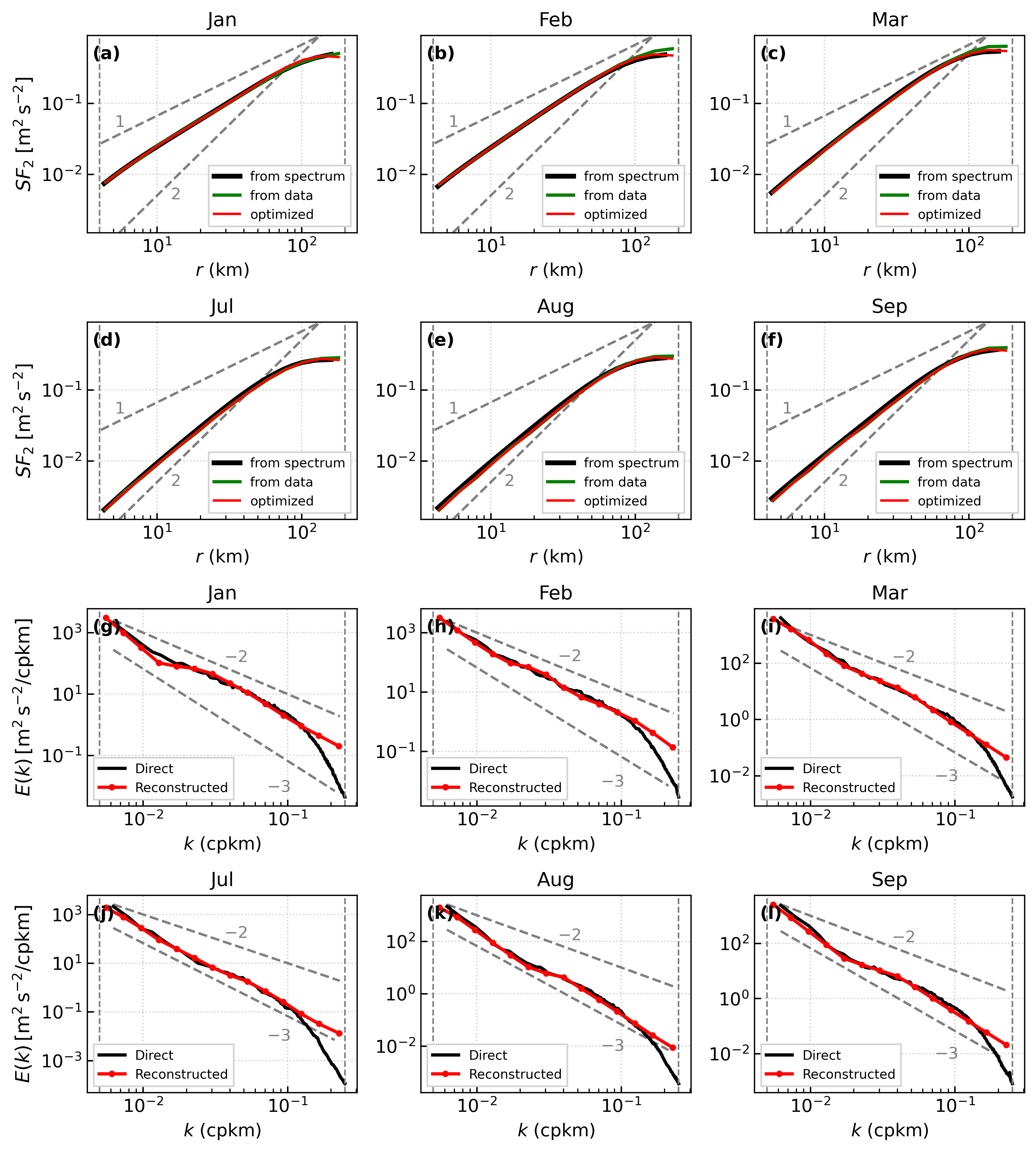}
 \caption{
Monthly averaged second-order structure functions, $SF_2(r)$, using MITgcm llc4320 model data, shown for January--March (a--c) and July--September (d--f). Black curves denote $SF_2(r)$ computed from the KE spectrum $E(k)$ of the model velocity fields (Eq.~\ref{eq:sf2_exact}), green curves are estimated directly from the model data, and red curves represent the optimized $SF_2(r)$ obtained using the regularized method; all three curves overlap closely. In panels (a--f), the vertical grey dashed lines mark the separation-distance range of $4-200$ km used in the analysis. The corresponding monthly KE spectra $E(k)$ are shown for January--March (g--i) and July--September (j--l), with black curves indicating spectra computed directly from the model velocity fields and red curves showing the reconstructed spectra from the regularized method. Vertical dashed lines in (g--l) mark the wavenumber range $0.005-0.25$ cpkm, corresponding to the separation scales used in (a--f).}
 \label{fig: mitgcm_plot}
\end{figure}

To estimate $SF_2(r)$, velocity pairs are obtained by randomly subsampling approximately $10\%$ of the grid points in the domain (about $4{,}000$ out of $40{,}000$). Computing all possible pairs from the full grid is computationally expensive, so we tested repeated subsampling and found that using $4{,}000$ points is sufficient for $SF_2(r)$ to converge. The resulting velocity pairs are grouped into logarithmically spaced bins according to their horizontal separation distance, with bin edges defined by a geometric progression. Within each bin, $r$ is defined as the mean separation distance of all pairs. For each pair, the velocity difference is computed and projected onto the longitudinal and transverse directions relative to $\mathbf{r}$, yielding $\delta u_L$ and $\delta u_T$ (Eq.~\ref{eq: sf2_LT}). The second-order structure function is then calculated as the bin-averaged sum of the squared longitudinal and transverse velocity differences (Eq.~\ref{eq:sf2_cal}). The lower bound of the separation $r$, marked by the vertical gray dashed lines in Figure~\ref{fig: mitgcm_plot}(a--f), is set to $4~\mathrm{km}$, which is approximately twice the model horizontal grid spacing, $\Delta x = 2~\mathrm{km}$, and corresponds to the minimum resolvable wavelength (Nyquist wavelength). The upper bound is constrained by the finite domain size, $L = 400~\mathrm{km}$; after windowing, the effective maximum separation is taken as $L/2 = 200~\mathrm{km}$.

The analysis domain is 400~km, so separations are meaningful only up to about half the domain size, which sets the 200-km limit of the fit. This choice reflects our focus on the mesoscale and submesoscale range: our primary interest is whether transitions in the kinetic energy spectral slope at these scales are captured in the reconstructed spectra, and the 200-km separation range is sufficient for this purpose. In addition, increasing the domain size would reduce the accuracy of the estimated $SF_2$. Since we subsample the model fields by retaining only 10\% of the grid points to make the pairwise calculations computationally feasible, a larger domain would distribute the same number of samples over a broader range of separations. This would reduce the number of pairs in each separation bin and increase the sampling error. Reducing this additional uncertainty would require applying statistical techniques, such as bootstrapping, to obtain more robust estimates of $SF_2$. To avoid this additional level of complexity while maintaining reliable $SF_2$ estimates, we limited our analysis to the chosen domain size.

The $E(k)$ is computed from the model velocity fields to provide a reference for comparison with the reconstructed spectra. The zonal and meridional velocity fields are first transformed into spectral space using the two-dimensional Fast Fourier Transform (FFT), yielding the zonal and meridional wavenumber components $k_x$ and $k_y$, respectively. The radial wavenumber is then calculated as $k=\sqrt{k_x^2+k_y^2}$, and $k$ space is partitioned into $N_{\mathrm{shell}}=N$ annular shells, where $N$ is the total number of discrete wavenumbers, thereby retaining the maximum wavenumber resolution without spectral smoothing. The 2D velocity power spectral density $\Phi$ is then computed from the squared magnitudes of the Fourier coefficients, and $E(k)$ is obtained by azimuthally averaging the 2D power over each annular shell and multiplying by the shell's central wavenumber following Eq.~\ref{eq: ek}.

\subsection{Reconstruction Performance for Model Data}

Figure~\ref{fig: mitgcm_plot}(a--f) shows close agreement between the optimized structure functions (red curves), the structure functions estimated directly from the model velocity fields (green curves), and the structure function computed from the KE spectra using Eq.~\ref{eq:sf2_exact} (black curves). It should be noted, that the separation grids differ slightly between the latter two estimates. The structure functions computed from the model KE spectra are evaluated on an \(r\) grid derived from the FFT wavenumber grid, whereas the structure functions estimated directly from the data are evaluated on an \(r\) grid defined by the separation bins. As a result, the directly estimated \(SF_2(r)\) extends slightly farther in \(r\) than the spectrum-based estimate.

\begin{figure}[htbp]
 \centering
 \noindent\includegraphics[width=25pc,angle=0]{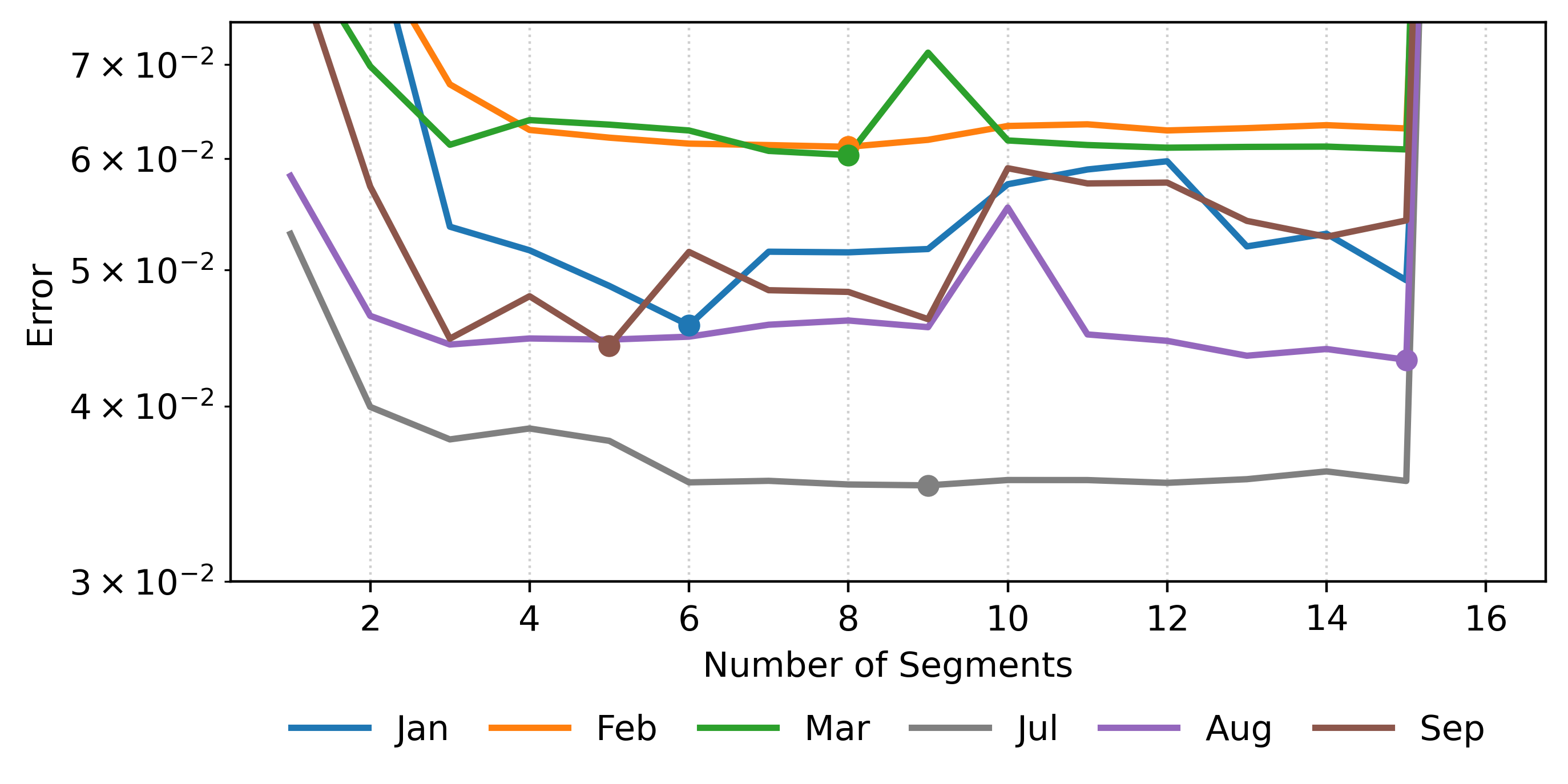}
 \caption{Error between the estimated and optimized second-order structure functions for MITgcm llc4320 model data case, defined in Eq.~\eqref{eq:error}, as a function of the number of spectral segments for different months. Solid curves show the error for each month as the number of segments varies. Dots indicate the optimal number of spectral segments selected for each month, corresponding to the minimum error (best-fit error) achieved by the fitting procedure.}
 \label{fig: err_mitgcm}
\end{figure}

Figure~\ref{fig: err_mitgcm} shows the error evaluated using Eq.~\eqref{eq:error} as a function of the number of spectral segments for each month. The directly estimated $SF_2(r)$ contains $N = 16$ data points for all months, and accordingly the optimization is performed by varying the number of segments from 1 to 16. For all months, the optimal number of spectral segments ranges between 5 and 10, with the exception of August, which requires 15 segments (Figure~\ref{fig: err_mitgcm}); the corresponding segment boundaries for the reconstructed KE spectra are shown in Supplemental Material Figure~S2. However, once the number of segments reaches $S=N$, the error increases sharply for all months, indicating instability in the overparameterized regime. This behavior is consistent with the idealized cases (Figure~\ref{fig: err_ideal}) and arises because the TRF algorithm becomes unstable when too many parameters are introduced, as discussed in the Appendix.

Figure~\ref{fig: mitgcm_plot}(g-l) shows the monthly reconstructed KE spectra (red curves) from the structure functions compared with the spectra calculated directly from the model velocity fields (black curves). For all months, the reconstructed spectra (red curves) closely follow the directly computed spectra (black curves) across all wavenumbers except the two highest-wavenumber points (near the Nyquist limit), indicating that the inversion 
robustly recovers the dominant spectral structure. The main departures occur at the two highest-wavenumber points near the Nyquist limit, because the method does not permit spectral slopes steeper than $-4$. During winter (Figure~\ref{fig: mitgcm_plot}(g--i)), both spectra exhibit an approximate $k^{-2}$ scaling for $10^{-2}$ cpkm $<k< 10^{-1}$ cpkm, with corresponding structure functions (Figure~\ref{fig: mitgcm_plot}(a--c)) scaling as $r^{1}$ ($10^{1}$ km $< r< 10^{2}$ km). The spectral and structure function slopes potentially indicate quasi-geostrophic dynamics modified by ageostrophic effects (e.g., frontogenesis) consistent with \cite{callies2015seasonality, rocha(2016mesoscale(a))}. In summer (Figure~\ref{fig: mitgcm_plot}(j--l)), the spectra are steeper, approaching $k^{-3}$ ($2\times 10^{-2}$ cpkm $<k< 10^{-1}$ cpkm), with structure functions (Figure~\ref{fig: mitgcm_plot}(d--f)) scaling as $r^{2}$ ($10$ km $<r< 50$ km), both slopes are indicative of interior, isotropic quasi-geostrophic turbulence \citep{rocha(2016mesoscale(a))}. Moreover, the directly computed $E(k)$ shows a small-wavenumber transition ($1.5 \times 10^{-2}$ to $3 \times 10^{-2}$ cpkm), most evident in January, August, and September, where the slope becomes steeper. This behavior is consistent with a shift in the governing dynamics and is clearly captured by the reconstruction, indicating that the method not only reproduces $SF_2(r)$ accurately but also resolves changes in spectral slope across scales.

\section{Application to drifter data}
\label{sec: Drifter_data}

\begin{figure}[htbp]
 \centering
 \noindent\includegraphics[width=35pc,angle=0]{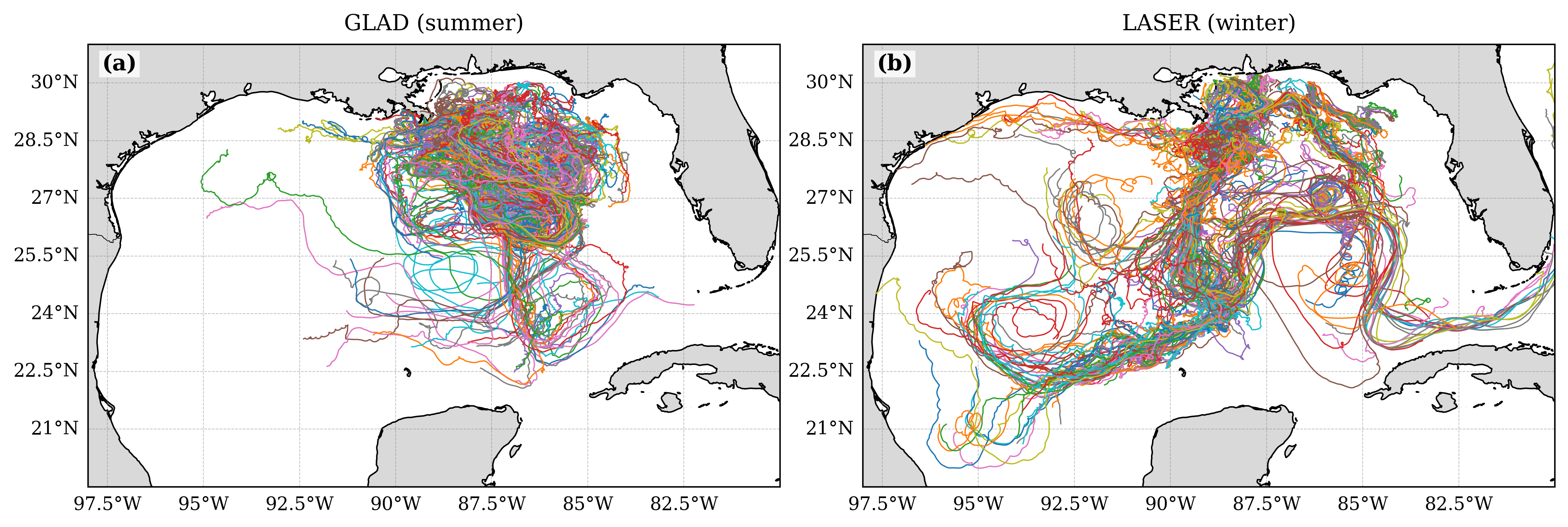}
 \caption{Surface drifter trajectories in the Gulf of Mexico from the GLAD and LASER experiments. (a) GLAD drifter trajectories during summer (July--August 2012) and (b) LASER drifter trajectories during winter (January--February 2016).}
 \label{fig: drifter_traj}
\end{figure}

After establishing confidence in our regularized method using synthetic spectra and high-resolution ocean model velocity fields, we apply it to sparsely sampled GLAD and LASER drifter data previously analyzed by \citet{balwada2016scale}, \citet{balwada2022}, and \citet{gutierrez2026improved}. The GLAD experiment was conducted during summer (July--August 2012) \citep{poje2014submesoscale}, while the LASER experiment took place during winter (January--February 2016) \citep{d2018ocean}; their trajectories are shown in Figure~\ref{fig: drifter_traj}. The GLAD and LASER experiments comprise 297 and 956 drifters, respectively, providing velocity measurements over spatial separations spanning $\mathcal{O}(10^{-3})$--$\mathcal{O}(10^{3})$~km and enabling sampling of submesoscale to mesoscale dynamics. In this study, we directly use the second-order structure functions estimates from \citet{balwada2022}, where the structure functions were calculated using drifters located north of $24^\circ$N, between $91^\circ$W and $84^\circ$W, and in ocean depths greater than 500~m to ensure near-homogeneity. 

For consistency with the idealized case study, $SF_2(r)$ is restricted to separations in the range $\mathcal{O}(10^{-1})$--$\mathcal{O}(10^{2})$~km, defining a truncated interval $[\tilde{r}_{\min},\,\tilde{r}_{\max}]$ contained within the full range $[r_{\min},\,r_{\max}]$ of the drifter dataset. Although $SF_2(r)$ is truncated at both ends, the wavenumber grid used for the construction is computed from the full $r$ range via Eq.~\ref{eq:rk_relation}, so that $k_{\min} = 2\pi / r_{\max}$ and $k_{\max} = 2\pi / r_{\min}$. This prevents the truncation from artificially narrowing the spectral range used in the forward model. The reconstructed spectrum is interpreted only over the wavenumber range $[\tilde{k}_{\min},\,\tilde{k}_{\max}] = [2\pi / \tilde{r}_{\max},\,2\pi / \tilde{r}_{\min}]$ corresponding to the retained separations, and only this range is shown in Figure~\ref{fig:drifter}.

\subsection{Spectral Estimation from Bootstrap Realizations}

In \citet{balwada2022}, the uncertainty in the $SF_2$ estimates is quantified using a modified block bootstrapping approach that accounts for the temporal and spatial correlations inherent in drifter pair samples. For each experiment, 600 bootstrap realizations of the $SF_2$ are generated by resampling $N^{\mathrm{DOF}}(r)$ (number of degrees of freedom) blocks with replacement from the concatenated set of drifter pair samples within each separation bin, where $N^{\mathrm{DOF}}(r)$ is the effective number of statistically independent velocity-difference samples at separation $r$. These 600 realizations serve as the ensemble of observed $SF_2$ curves to which our regularized method is applied. Further details of the structure-function calculation and uncertainty estimation are provided in \citet{balwada2022}. 

\begin{figure}[t]
 \centering
 \noindent\includegraphics[width=38pc,angle=0]{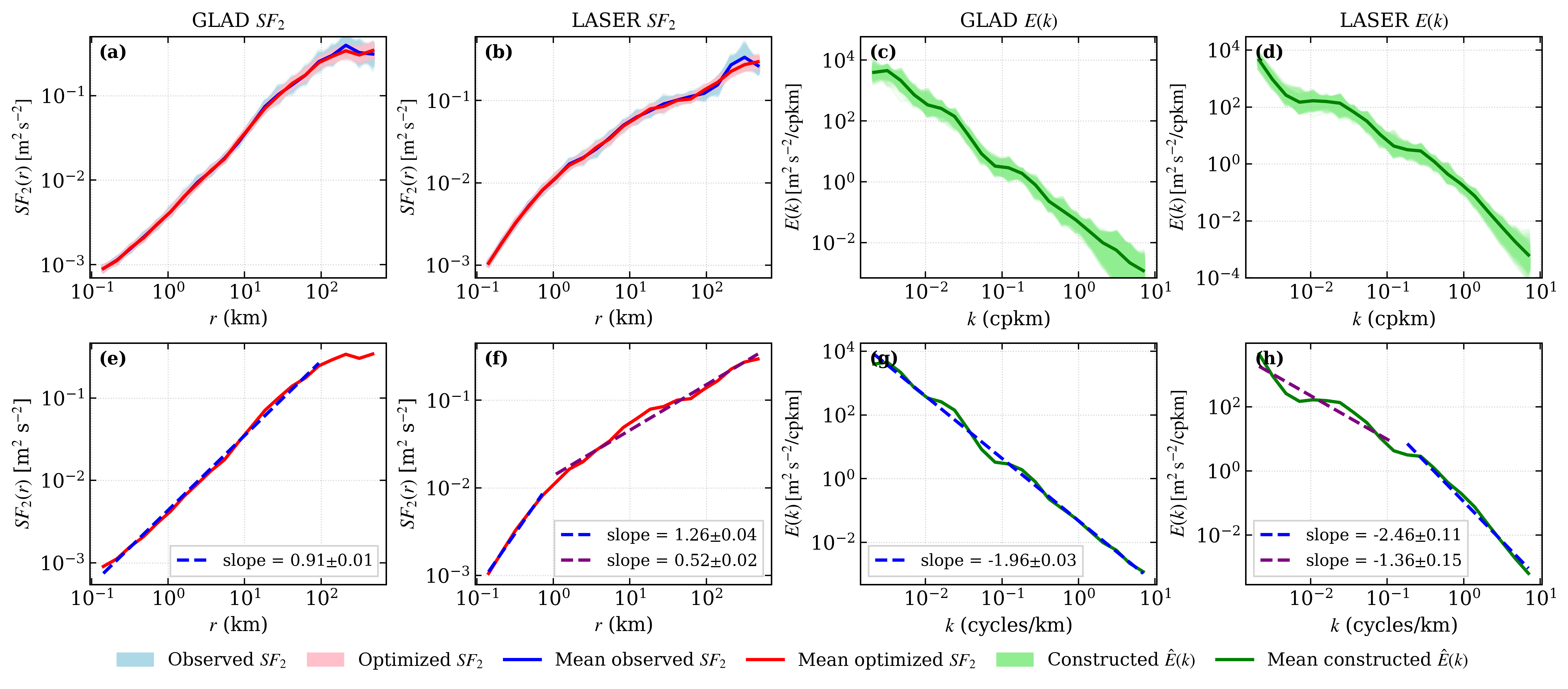}
 \caption{Results from the application of the regularized fitting method to the GLAD and LASER drifter datasets. (a)--(b) Ensemble of observed $SF_2$ (blue shading) and optimized $SF_2$ (pink shading) across all $600$ bootstrap realizations, with the mean observed (blue) and mean optimized (red) curves overlaid. (c)--(d) Corresponding ensemble of constructed $E(k)$ (lightgreen shading) with the mean constructed $E(k)$ (dark green). (e)--(f) Mean observed $SF_2$  (blue) and mean optimized $SF_2$  (red) with power-law fits (dashed lines) in log--log space. (g)--(h) Corresponding power-law fits to the mean constructed $E(k)$}
 % \caption{Enter the caption for your figure here.  Repeat as
 % necessary for each of your figures. Figure from \protect\cite{Knutti2008}.}
 \label{fig:drifter}
\end{figure}

The 600 bootstrap realizations of $SF_2$ for the GLAD and LASER experiments are shown as the light blue spread in Figure~\ref{fig:drifter}, with the corresponding optimized $SF_2$ realizations shown in pink. The mean optimized $SF_2$ (red curve) closely follows the mean observed $SF_2$ (blue curve) across most separations, with deviations becoming more apparent at larger separations, consistent with the behavior in Figure~\ref{fig: sf2_ensemble}(a--c) and (g--i). This reduced agreement at large $r$ reflects the monotonic decrease in degrees of freedom with separation (Supplementary Materials Figure~S4a of \citet{balwada2022}), which increases the uncertainty in the observed $SF_2$ and makes the fitting less constrained at larger scales.

Supplemental Material Figure~S3 shows the best-fit error and the optimal number of spectral segments for each of the 600 bootstrap realizations of the GLAD and LASER datasets. The GLAD (blue) and LASER (green) cases have mean best-fit errors of $2 \times 10^{-3}$ and $3 \times 10^{-3}$, respectively. The larger errors in LASER are likely associated with its lower effective degrees of freedom in the $SF_2$ estimates. Many LASER drifters lost their drogues early in the deployment, reducing the number of valid trajectory pairs and shortening the effective sampling period to about 60 days, compared to 90 days for GLAD \citep{balwada2022}. As a result, the LASER-based $SF_2$ estimates are noisier, especially at larger separations where the number of independent samples is already limited. Even so, the best-fit errors for both datasets remain on the order of $10^{-3}$ and are tightly clustered, indicating that the least-squares fitting produces stable agreement between the optimized and observed $SF_2(r)$ despite sampling variability.

The corresponding ensemble of constructed KE spectra is shown in Figure~\ref{fig:drifter}(c,d), where the green shading indicates the spread across the 600 realizations and the dark green curve denotes the ensemble-mean constructed spectrum. At low wavenumbers, the ensemble-mean $E(k)$ becomes less smooth, particularly for LASER, reflecting the propagation of the larger $SF_2$ uncertainties at large separations into the constructed spectrum at low $k$.

\subsection{Estimated Spectra and Scaling Behavior}

Since no direct independent estimate of $E(k)$ is available from these drifter observations, the performance of the method is assessed by examining whether the expected power-law relationship between the estimated mean $SF_2(r)$ and the constructed mean $E(k)$ is satisfied, as mentioned in Section \ref{sec:method}. The slopes are estimated by fitting a linear line in log–log space using a least-squares approach. The associated uncertainties are quantified from the standard error of the fit, which reflects the scatter of the data about the fitted line and provides a measure of the confidence in the estimated slope. Figure \ref{fig:drifter}(e--f) shows the fitted slopes for $SF_2(r)$, while \ref{fig:drifter}(g--h) show the corresponding fits for $E(k)$. For the GLAD dataset, a single scaling regime is identified, yielding slopes of $0.91 \pm 0.01$ for $SF_2(r)$ and $-1.96 \pm 0.03$ for $E(k)$. In contrast, the LASER dataset exhibits a transition in scaling, and two linear fits are applied, resulting in $SF_2(r)$ slopes of $1.26 \pm 0.04$ and $0.52 \pm 0.02$, with corresponding spectral slopes of $-2.46 \pm 0.11$ and $-1.36 \pm 0.15$. In all cases, the estimated slopes satisfy the expected power-law relationship between $SF_2(r)$ and $E(k)$ discussed in Section~\ref{sec: Two-point statistics}, supporting the consistency of the method. The recovered slope values are in close agreement with those reported in \cite{qian2025inferring}, although the slopes in that study were determined visually rather than through a formal fitting procedure.

\section{Discussion and Conclusion}
\label{sec: discussion}

We present a regularized framework to estimate KE spectra from second-order structure functions, enabling spectral inference from sparsely sampled Lagrangian observations. By avoiding direct inversion of the Hankel-transform relationship, which is ill-conditioned because the oscillatory Bessel kernel makes the inversion highly sensitive to noise, finite sampling, and the limited range of observed separations (Section~2 and Supplemental Material Figure~S1), the method instead relies on forward modeling and constrained least-squares fitting in physical space. This approach provides a flexible representation of the spectrum that accommodates multiple scaling regimes without imposing a single power-law form.

The framework was evaluated against three reference settings of increasing complexity: idealized analytic spectra, the MITgcm llc4320 simulation, and surface drifter observations from the GLAD and LASER experiments in the Gulf of Mexico. The estimated $E(k)$ shows close agreement with (i)~the prescribed analytic spectra in the idealized tests, (ii)~the spectra computed directly from the MITgcm velocity fields, and (iii)~the $SF_2$--$E(k)$ power-law scaling expected from theory in the drifter application, demonstrating that the method preserves the physical correspondence between $SF_2(r)$ and $E(k)$ over the wavenumber range corresponding to the sampled separations. The method also captures the distinct spectral slopes and partially capture slope transitions associated with different theoretical regimes, such as quasi-geostrophic, surface quasi-geostrophic, and submesoscale dynamics.

Although the method robustly captures the overall spectral structure, the analysis highlights two key practical constraints. First, uncertainties in $SF_2(r)$, particularly at large separations, propagate into the estimated spectra and limit accuracy at low wavenumbers. Second, noise in $SF_2(r)$ also degrades the precise localization of scale transitions: under noisy conditions (Figure~\ref{fig: sf2_ensemble}), the wavenumber at which the spectral slope changes is recovered only to within a small range, even though the surrounding slopes themselves remain well constrained. Both limitations reflect the quality and sampling of the data rather than the inversion framework itself. It is worth noting that the slope bounds of $k^{-4}$ and $k^{1}$ used here are a choice of the present implementation rather than an inherent limitation of the method, which can in principle accommodate any slope range. These particular bounds were selected because they encompass the physically relevant regimes from steep geostrophic turbulence to shallow submesoscale and wave-dominated dynamics; they can be widened or narrowed to suit a different application without any change to the underlying framework. In practice, however, the bounds also play a regularizing role: allowing too much flexibility in the slope range tends to yield physically unrealistic $E(k)$ estimates, because the optimizer can exploit unconstrained slope values to overfit local features of $SF_2(r)$. The chosen bounds therefore reflect a balance between physical generality and the smoothness of the recovered spectrum.

Future extensions of this work fall into two broad categories: applying the framework to a wider range of flow fields and observational datasets; and refining the inversion procedure itself to enhance its flexibility and robustness. The framework can be applied directly to a wide range of existing datasets and flow partitions. One particularly natural extension is to decomposed flow fields obtained through wave--vortex or Helmholtz decompositions \citep{buhler2014wave}, in which the rotational and divergent components are inverted separately to characterize the distinct spectral signatures of balanced geostrophic motion and unbalanced internal-wave and submesoscale dynamics. 

Beyond these direct applications, several refinements to the framework itself merit future development. In the current formulation, $E(k)$ is represented by piecewise power-law segments joined through Heaviside functions, which are not differentiable and impose sharp (though continuous) transitions between segments. Replacing this discretization with smoother, differentiable basis functions would yield more continuous spectral representations across scales and may improve the conditioning of the inversion. Another limitation is that the subrange boundaries are fixed a priori, uniformly in $\log k$, so a slope transition in the true spectrum can be located only to within the width of a subrange. Treating the transition wavenumbers as free parameters of the inversion --- at the cost of a nonlinear and possibly less stable optimization --- could sharpen the localization of regime transitions, and is a natural refinement for future work. Moreover, this representation is designed for smoothly varying spectra and does not explicitly account for narrowband spectral features or spectral lines. Extending the framework to include such localized spectral features would require additional basis functions or constraints that allow departures from a smoothly varying background spectrum. Irregular sampling of velocity observations can introduce scale-dependent uncertainty in \(SF_2\) because the number and distribution of point pairs vary across separation ranges. Statistical approaches, such as bootstrapping, are commonly used to quantify and reduce this uncertainty in \(SF_2\) estimates. Incorporating such uncertainty quantification, together with noise-aware or weighted fitting strategies that explicitly account for the scale-dependent uncertainty in $\text{SF}_2(r)$ would further enhance performance, especially at large separations where the number of independent pairs is limited. Together, these developments would extend the applicability of the framework to still noisier and more sparsely sampled observations while preserving its current flexibility.

% \appendix

% \section{Error spike due to Jacobian rank-deficit}
% \label{appendix:jacobian}

\clearpage
%%%%%%%%%%%%%%%%%%%%%%%%%%%%%%%%%%%%%%%%%%%%%%%%%%%%%%%%%%%%%%%%%%%%%
% ACKNOWLEDGMENTS
%%%%%%%%%%%%%%%%%%%%%%%%%%%%%%%%%%%%%%%%%%%%%%%%%%%%%%%%%%%%%%%%%%%%%
\acknowledgments
This work was funded by NSF Physical Oceanography grant number OCE- 2242109.

%%%%%%%%%%%%%%%%%%%%%%%%%%%%%%%%%%%%%%%%%%%%%%%%%%%%%%%%%%%%%%%%%%%%%
% DATA AVAILABILITY STATEMENT
%%%%%%%%%%%%%%%%%%%%%%%%%%%%%%%%%%%%%%%%%%%%%%%%%%%%%%%%%%%%%%%%%%%%%
% 
%
\datastatement 
The MITgcm LLC4320 data can be accessed using the code available at \url{https://doi.org/10.5281/zenodo.11373789}. The surface drifter data are available at \url{https://data.gulfresearchinitiative.org/}. The code for the method and figure generation is available at \url{https://doi.org/10.5281/zenodo.19704378}.

%%%%%%%%%%%%%%%%%%%%%%%%%%%%%%%%%%%%%%%%%%%%%%%%%%%%%%%%%%%%%%%%%%%%%
% APPENDIXES
%%%%%%%%%%%%%%%%%%%%%%%%%%%%%%%%%%%%%%%%%%%%%%%%%%%%%%%%%%%%%%%%%%%%%
%
%% If only one appendix, use

\appendix

\section{Error spike due to Jacobian rank-deficit}
\label{appendix:jacobian}

Figures \ref{fig: err_ideal} and \ref{fig: err_mitgcm} show that the error spike occurs when the number of spectral segments equals the total number of data points, $N$. The non-linear least-squares fitting in this study is performed using the Trust Region Reflective (TRF) algorithm, which operates on the Jacobian matrix $J \in \mathbb{R}^{N \times (S+1)}$, where $J_{ij} = \partial r_i / \partial p_j$ quantifies the sensitivity of the $i$th residual to the $j$th parameter. Since $\mathrm{rank}(J) \leq \min(N, S+1)$, three regimes arise. For $S < N-1$, the system is overdetermined; the Jacobian $J$ can attain full column rank. At $S = N-1$, the system is exactly determined;  $J$ becomes square and remains full rank. In these cases, each parameter is, in principle, uniquely identifiable from the residuals. For $S \geq N$, $J$ becomes fat and the system is under-determined, so full column rank is no longer possible (rank-deficit) and at least one parameter direction is left unconstrained by the data.

We confirmed this behavior for the idealized ocean case. At $S = 28$, $J \in \mathbb{R}^{30 \times 29}$ has full column rank ($\mathrm{rank}(J) = 29$) with no null space. At $S = 29$, $J$ becomes square ($30 \times 30$) while remaining full rank, indicating a well-posed system. In contrast, at $S = 30$, $J$ becomes a fat matrix ($30 \times 31$), and the rank--nullity theorem guarantees a nontrivial null space,
\begin{equation}
(S + 1) - \mathrm{rank}(J) \geq 1,
\label{eq:nullspace}
\end{equation}
independent of the specific problem structure. The singular value decomposition of $J$ at $S = 30$ confirms the presence of atleast one singular value to be zero, making the system ill-conditioned.

%% If more than one appendix, use \appendix[<letter>], e.g.,

%\appendix[A] 

%% Appendix title is necessary! For appendix title:

%\appendixtitle{Title of Appendix}

%%% Appendix section numbering (note, skip \section and begin with \subsection)
%
% \subsection{First primary heading}

% \subsubsection{First secondary heading}

% \paragraph{First tertiary heading}

%%%%%%%%%%%%%%%%%%%%%%%%%%%%%%%%%%%%%%%%%%%%%%%%%%%%%%%%%%%%%%%%%%%%%
% REFERENCES
%%%%%%%%%%%%%%%%%%%%%%%%%%%%%%%%%%%%%%%%%%%%%%%%%%%%%%%%%%%%%%%%%%%%%
% Make your BibTeX bibliography by using these commands:
\bibliographystyle{ametsocV6}
\bibliography{references_modified}
\end{document}

\maketitle

\section{Derivation of KE Spectrum from the Second-Order Structure Function}

The forward relation between the second-order structure function, $SF_2(r)$, and the kinetic energy spectrum, $E(k)$, is
\begin{equation}
    SF_2(r) = 2\int_0^\infty E(k)\left[1 - J_0(kr)\right]\,dk,
    \label{eq:sf2}
\end{equation}
where $J_0$ is the zeroth-order Bessel function of the first kind. In the limit $r \to \infty$, $J_0(kr) \to 0$, so Eq. \ref{eq:sf2} reduces to,
\begin{equation}
    SF_2(\infty) = 2\int_0^\infty E(k)\,dk.
\end{equation}

Subtracting $SF_2(r)$ from $SF_2(\infty)$ and dividing by 2 gives,
\begin{equation}
\frac{SF_2(\infty) - SF_2(r)}{2}
= \int_0^\infty E(k)\,J_0(kr)\,dk
= \int_0^\infty \frac{E(k)}{k}\,J_0(kr)\,k\,dk,
\label{eq:hankel}
\end{equation}
which is the zeroth-order Hankel transform, $\mathcal{H}_0$, of $E(k)/k$:
\begin{equation}
    \mathcal{H}_0\left\{\frac{E(k)}{k}\right\}(r)
    = \frac{SF_2(\infty) - SF_2(r)}{2}.
\end{equation}

Because the zeroth-order Hankel transform is self-reciprocal, applying it to both sides yields
\begin{equation}
    \frac{E(k)}{k}
    = \int_0^\infty \frac{SF_2(\infty) - SF_2(r)}{2}\,J_0(kr)\,r\,dr.
\end{equation}
Multiplying both sides by $k$ gives
\begin{equation}
    E(k)
    = \int_0^\infty \frac{SF_2(\infty) - SF_2(r)}{2}\,J_0(kr)\,k r\,dr,
    \label{eq:ek_inversion}
\end{equation}
which provides the inverse relation for recovering the kinetic energy spectrum from the second-order structure function. 
\begin{figure}[htbp]
 \centering
 \includegraphics[width=38pc]{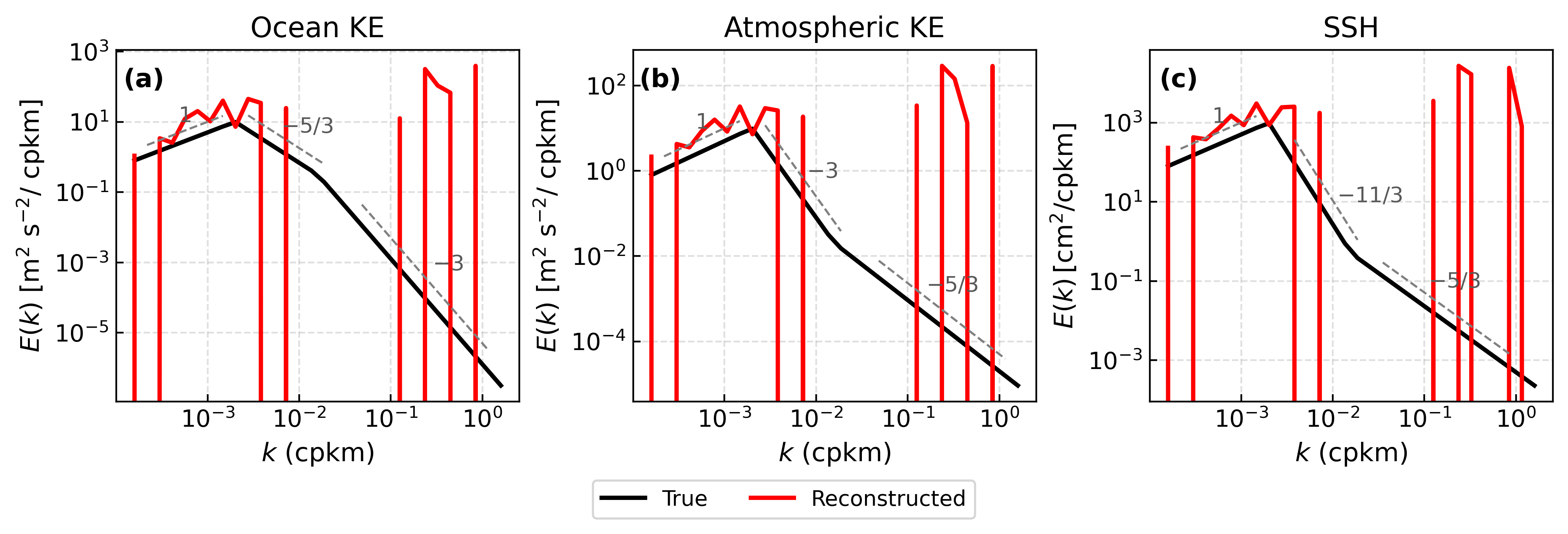}
 \caption{Comparison of the true and reconstructed kinetic energy spectra, $E(k)$, for three idealized cases: (a) ocean KE, (b) atmosphere KE, and (c) SSH. The black curves denote the true spectra, and the red curves denote the spectra reconstructed from the corresponding second-order structure functions, using the inverse relation given by Eq.~\ref{eq:ek_inversion}.}. 
 \label{fig: ideal_inversion}
\end{figure}

Figure~\ref{fig: ideal_inversion} shows that directly reconstructing \(E(k)\) from \(SF_2(r)\) using Eq.~\ref{eq:ek_inversion} is numerically unstable in all three idealized cases: ocean KE, atmospheric KE, and SSH. In each panel, the reconstructed spectra (red curves) deviate strongly from the prescribed true spectra (black curves) and exhibit large oscillations and spurious spikes across the wavenumber range. This behavior reflects the ill-conditioned nature of the direct inverse transform: the inversion relies on \(SF_2(r)\) over a finite range of separations and involves an oscillatory Bessel kernel, both of which amplify numerical errors and make the recovered spectrum highly sensitive to discretization. As a result, the direct inverse relation does not provide a robust estimate of \(E(k)\), even for simple idealized spectra, thereby motivating the use of the regularized fitting approach.

\section{Piecewise Power-Law Representation of the Reconstructed Model Spectra}

\begin{figure}[htbp]
 \centering
 \includegraphics[width=30pc]{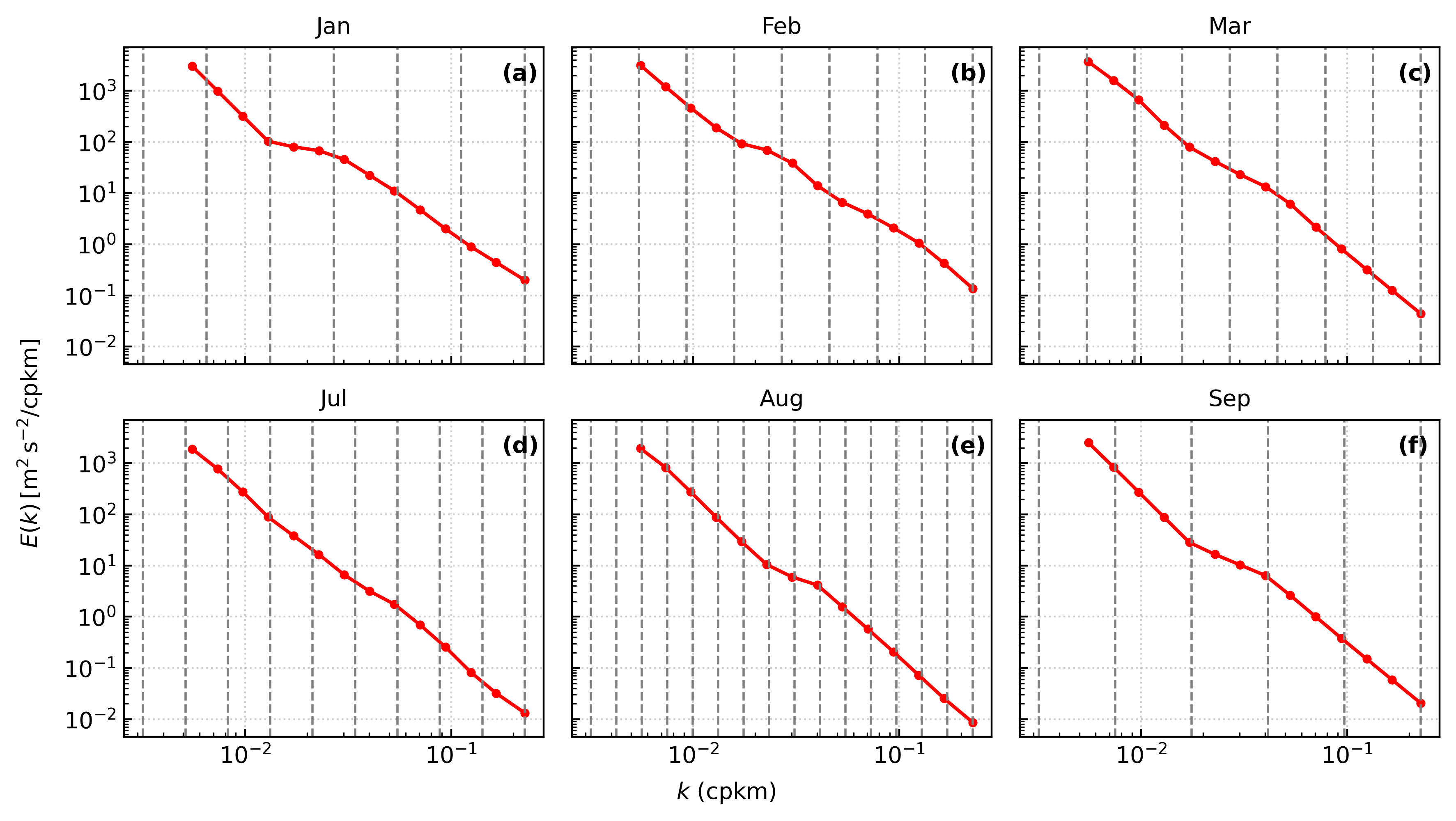}
 \caption{Reconstructed kinetic energy spectra for January–March (a–c) and July–September (d–f) using MITgcm llc4320 model output, also shown in red curves in Fig. 5g-l. Gray dashed vertical lines denote the segment boundaries, which are uniformly spaced in logarithmic wavenumber space and define the piecewise power-law representation.}
 \label{fig: model_seg}
\end{figure}

Figure~\ref{fig: model_seg} shows the reconstructed kinetic energy spectra for the MITgcm llc4320 model for representative winter and summer months, with gray dashed vertical lines indicating the segment boundaries, which are uniformly spaced in logarithmic wavenumber space. The discrete wavenumber points used in the reconstruction fall either within a segment or on its boundaries, as determined by the algorithm.

\section{Best-fit Error Distribution for GLAD and LASER $SF_2$}
\begin{figure}[htbp]
 \centering
 \includegraphics[width=30pc]{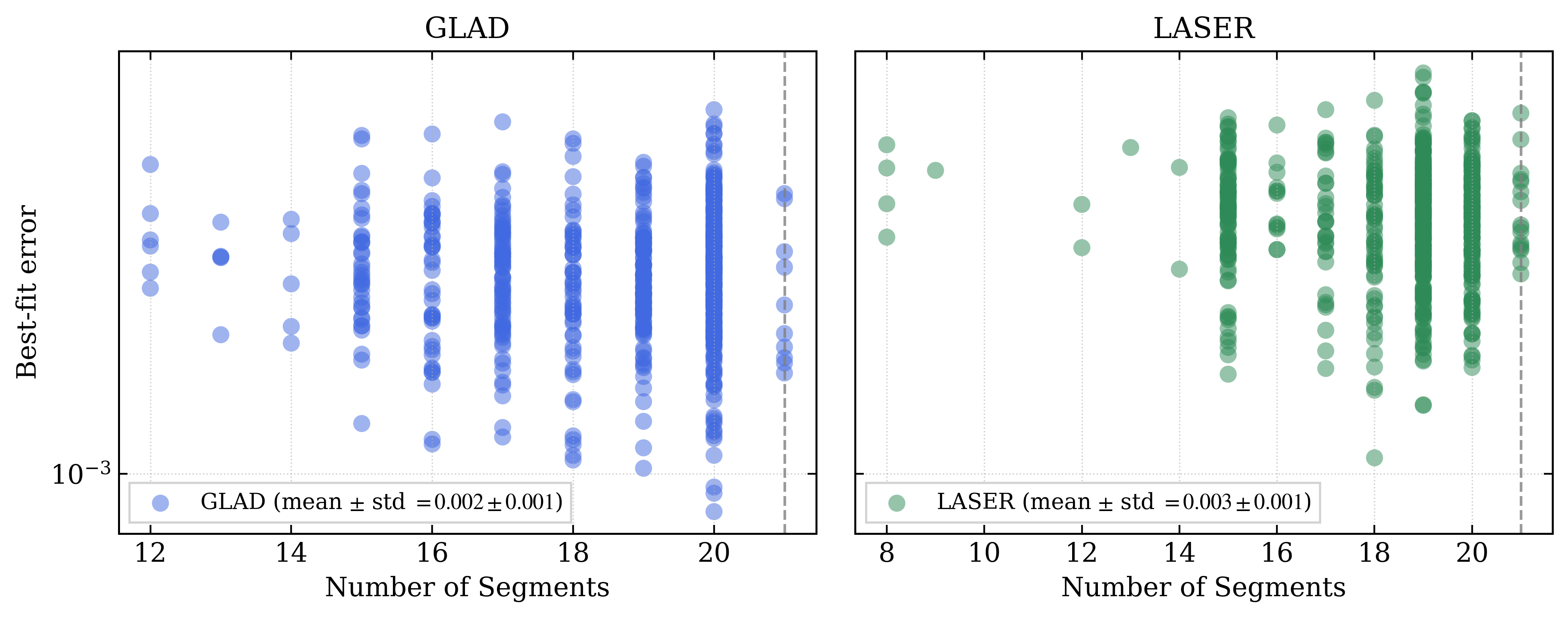}
 \caption{Best-fit error between the observed and optimized second-order structure functions as a function of the number of spectral segments for the GLAD (left) and LASER (right) drifter datasets, based on 600 realizations each. The vertical gray dashed line marks the maximum number of spectral segments (21), corresponding to the number of discrete separation points used in the fitting.}
 \label{fig: drifter_error}
\end{figure}

Figure~\ref{fig: drifter_error} shows the distribution of best-fit errors and the optimal number of spectral segments for each of the 600 bootstrap realizations of the GLAD (left) and LASER (right) datasets. For both datasets, $SF_2$ is evaluated at $N = 21$ discrete separation distances, so the optimization is performed by varying the number of segments from 1 to 21, with the gray vertical dashed line indicating $S = 21$. Each point represents the optimal number of segments and corresponding best-fit error for an individual realization. 

\section{Sensitivity of the Spectral Inversion to Non-Power-Law Spectra}

\begin{figure}[htbp]
\centering
\includegraphics[width=0.7\linewidth]{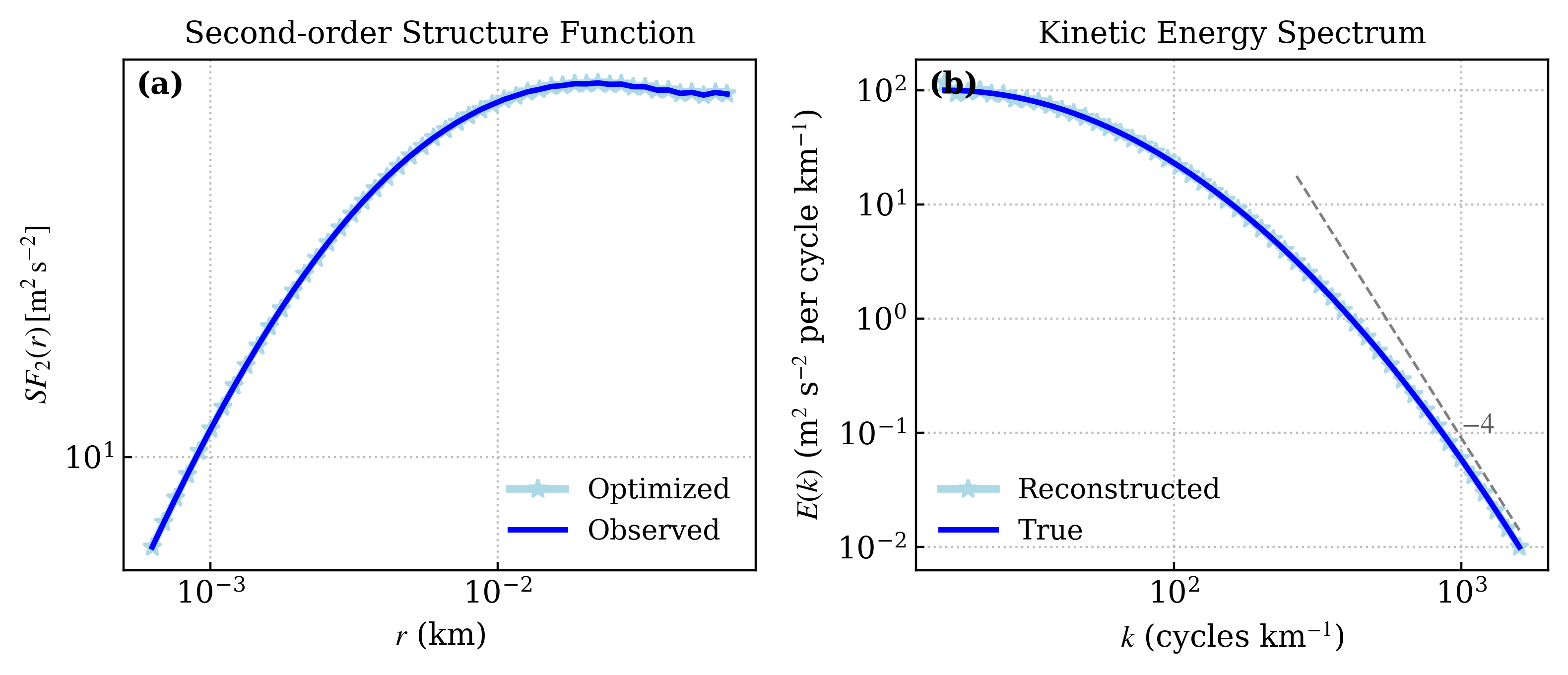}
\caption{(a) Prescribed (blue) and optimized (light blue) second-order structure functions($SF_2$). (b) True (blue) and reconstructed (lightblue) kinetic energy spectra.}
\label{fig:parabolic_spectrum}
\end{figure}

To evaluate the method for spectra that deviate from a simple power law, we construct a synthetic wavenumber spectrum $E(k)$ with a spectral slope that progressively steepens with increasing wavenumber $k$. This is achieved by allowing the power-law exponent to vary logarithmically with $k$. The spectrum is defined as,

\begin{equation}
E(k) = E_0 \left( \frac{k}{k_0} \right)^{-\left[2 + \log_{10}(k/k_0)\right],
\label{eq:parabolic}
}\end{equation}

where $E_0$ is a reference energy amplitude and $k_0$ is a
reference wavenumber. The resulting spectrum is shown in Figure
\ref{fig:parabolic_spectrum}b (blue). The corresponding \(SF_2\), computed using Eq.~\ref{eq:sf2}, is shown in Fig.~\ref{fig:parabolic_spectrum}a (blue) and is used to reconstruct the KE spectrum.

The optimized $SF_2$ (light blue) closely reproduces the prescribed $SF_2(r)$ (blue) across the full range of separation distances (Fig. \ref{fig:parabolic_spectrum}a). The corresponding reconstructed KE spectrum (light blue) agrees closely with the true spectrum (blue), capturing both the shallower low-$k$ range and the steep high-$k$ roll-off, where the local slope approaches $-4$ (dashed line; Figure \ref{fig:parabolic_spectrum}b). We choose a slope of $-4$ because it corresponds to the lower bound of the slope parameter $\alpha$ permitted by our method. This agreement demonstrates that our method can accurately recover $E(k)$ even when the underlying spectrum departs substantially from a single power law.

\section{Sensitivity of the Structure Function to the Number of Sampling Points}

\begin{figure}[htbp]
\centering
\includegraphics[width=0.6\linewidth]{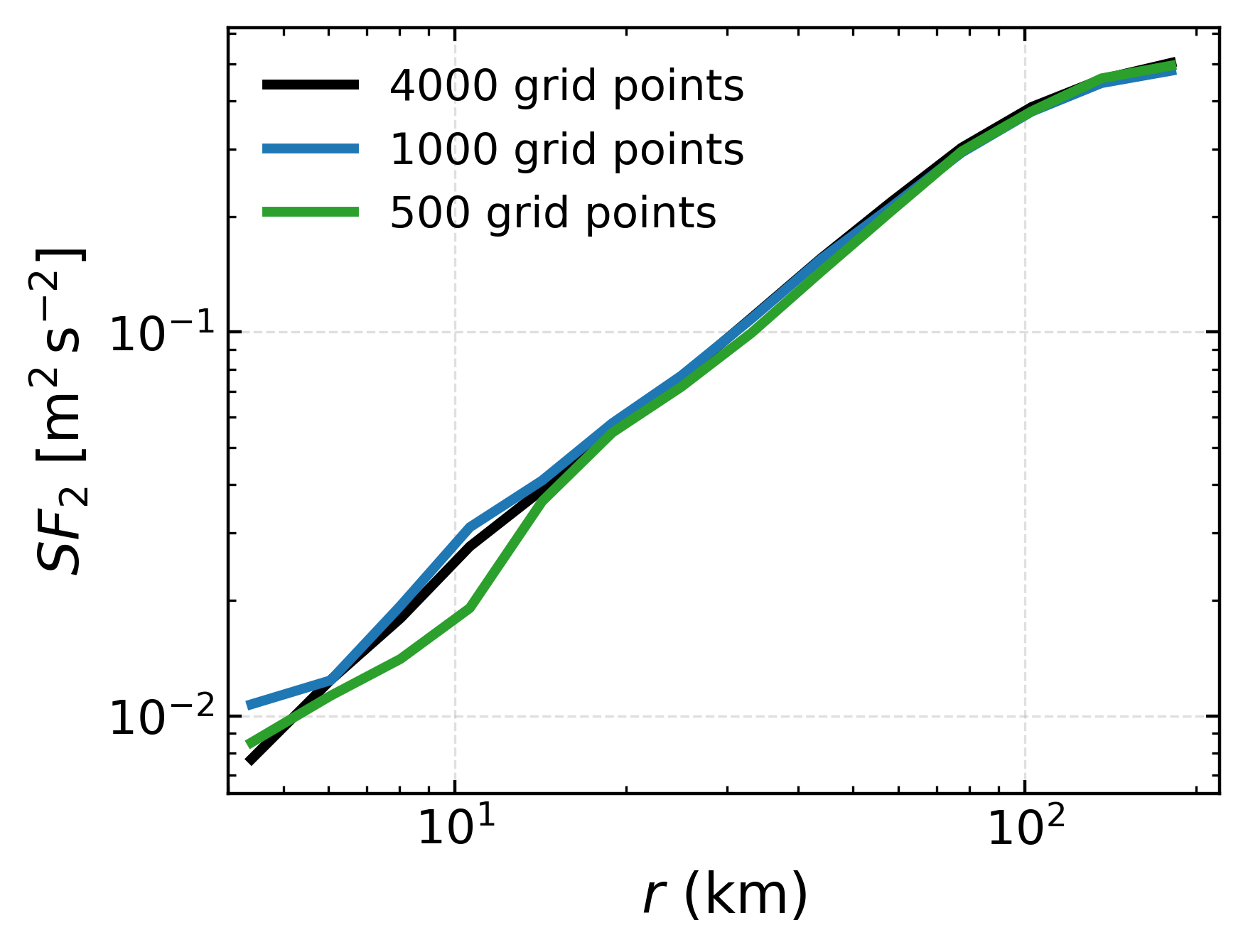}
\caption{Second-order structure function ($SF_2$) computed from the LLC4320 velocity
fields using 4000 (black), 1000 (blue), and 500 (green) randomly selected grid
points. The estimates agree closely for $r \gtrsim 15$~km; deviations at
smaller separations reflect the reduced number of point pairs per bin.
}
\label{fig:sf2_subsampling}
\end{figure}

We repeated the $SF_2$ calculation for the MITgcm llc4320 case using randomly selected 500, 1000 and 4000 grid points, and compared the results. Figure \ref{fig:sf2_subsampling} shows the three estimates. Over most of the resolved separation range (\(r \gtrsim 15~\mathrm{km}\)), the \(SF_2\) estimates obtained using 500, 1000, and 4000 sampled grid points are nearly indistinguishable. Differences are confined primarily to the smallest separations. For 500 sampled points, the \(SF_2\) is underestimated at \(r \lesssim 15~\mathrm{km}\), while the 1000-point estimate shows a slight increase in the smallest separation bins.

These results indicate that the $SF_2$ estimate is robust to a substantial
reduction in the number of sampling points: a deployment of order 1000
instruments would be sufficient to estimate $SF_2$ over the separation range analyzed here, with degradation confined to the smallest separations. We note, however, that this test retains the idealized sampling of the model analysis (points distributed over the full domain at fixed positions); realistic drifter deployments have smaller initial spans and evolving, inhomogeneous pair distributions, and a more comprehensive assessment using virtual drifter trajectories is therefore needed.

